\documentclass[12pt,twoside]{article}
\usepackage[eqsecnum]{cmp2e}
\usepackage{amsbsy,amssymb}
\usepackage{epsf}

\title[Thermodynamics of a pseudospin-electron model without correlations]%
 {Thermodynamics of a pseudospin-\\ electron model without correlations}
 \address{Institute for Condensed Matter Physics\\
 of the National Academy of Sciences of Ukraine,\\
 1 Svientsitskii Str., 79011 Lviv, Ukraine}
\author{I.V.Stasyuk, A.M.Shvaika, K.V.Tabunshchyk}
\date{Received February 1, 1999}

\begin{document}
\maketitle \setcounter{page}{109}

\newcommand{\ff}{\hspace*{-1pc}}
\newcommand{\fl}{\hspace*{-6pc}}
\newcommand{\fll}{\hspace*{-4pc}}
\newcommand{\rs}{\rm\scriptscriptstyle}
\newcommand{\bk}{\boldsymbol {k}}
\newcommand{\bq}{\boldsymbol {q}}
\begin{abstract}
{Thermodynamics of a pseudospin-electron model
without correlations is investigated.
The correlation functions, the mean values of pseudospin
and particle number, as well as the thermodynamic potential are calculated.
The calculation is performed by a diagrammatic method in the mean field
approximation. Single-particle Green functions are taken in the
Hubbard-I approximation. The numerical research
shows that an interaction between the electron
and pseudospin subsystems leads in the $\mu={}$const regime to the
possibility of the first order phase transition at the temperature change
with the jump of the pseudospin mean value $\langle S^z\rangle$
and reconstruction of the electron spectrum. In the regime $n={}$const, an
instability with respect to phase separation in the electron subsystem can
take place for certain values of the model parameters.}
\end{abstract}

\section{Introduction}

The model considering an interaction of electrons with a local anharmonic
mode of lattice vibrations has been used in the recent years in the theory
of high-tem\-pe\-ra\-tu\-re
superconducting crystals. Particularly, such a property
is characteristic of the vibrations of the so-called apex oxygen ions
O$_{\rm IV}$ along the $c$-axis direction of the layered compounds of the
YBa$_2$Cu$_3$O$_7$-type structure \cite{1,2,3}. An important role of
the apex oxygen and its anharmonic vibrations in the phase transition into
the superconducting state has already been mentioned \cite{4,5} and a
possible connection between the superconductivity and lattice instability of
the ferroelectric type in high-$T_{\rm c}$
superconducting compounds has been
discussed \cite{6,7}. In the case of a local double-well potential, the
vibrational degrees of freedom can be presented by pseudospin variables.
The Hamiltonian of the derived in this way pseudospin-electron model has
the following form~\cite{8}:
\begin{equation}
\label{1.1}
H=\sum_iH_i+\sum_{ij\sigma}t_{ij}b^+_{i\sigma}b_{j\sigma},
\end{equation}
and includes, besides the terms describing electron transfer ($\sim t_{ij}$),
the electron correlation ($U$-term), interaction with the anharmonic mode
($g$-term), the energy of the tunnelling splitting ($\Omega$-term)
and the energy of the anharmonic potential asymmetry ($h$-term) in the
single-site part
\begin{equation}
\label{1.2}
H_i=Un_{i\uparrow}n_{i\downarrow}+E_0(n_{i\uparrow}+n_{i\downarrow})+
g(n_{i\uparrow}+n_{i\downarrow})S^z_i-\Omega S^x_i
-hS^z_i.
\end{equation}
Here, $E_0$ gives the origin for energies of the electron states at
the lattice site ($E_0=-\mu$).

In this paper,
our aim is to obtain expressions for the correlation
functions which determine dielectric susceptibility
and the mean values of
pseudospin and particle number operators, as well as the thermodynamic
potential in the case of $\Omega =0$ and the absence of the
Hubbard correlation $U=0$.

We perform the numerical calculations and investigate the mean
values of pseudospin and particle number operators with
the change of asymmetry parameter $h$
($T$=const) or temperature $T$
($h$=const) for the cases of a fixed chemical potential value
(regime $\mu$=const) and a constant mean particle number (regime
$n$=const). An analysis of the thermodynamic properties of the
pseudospin-electron model in the case of absence of the electron
correlation is also made.

\section{Hamiltonian and initial relations}

We shall write the Hamiltonian of the model and the operators which
correspond to physical quantities in the second quantized form using
operators of the electron creation (annihilation) at a site with a
certain pseudospin orientation
\begin{eqnarray}
\label{2.1}
 a_{i\sigma}=b_{i\sigma}\big(
 \displaystyle
 \frac{1}{2}+S^z_i\big),\qquad &
 a^+_{i\sigma}=b^+_{i\sigma}\big(
 \displaystyle
 \frac{1}{2}+S^z_i\big),\nonumber \\
 \tilde{a}_{i\sigma}=b_{i\sigma}\big(
 \displaystyle
 \frac{1}{2}-S^z_i\big),\qquad &
 \tilde{a}^+_{i\sigma}=b^+_{i\sigma}\big(
 \displaystyle
 \frac{1}{2}-S^z_i\big).
\end{eqnarray}
Then, we obtain the following expression for the initial Hamiltonian:
\begin{eqnarray}
\label{2.2}
 H&=&H_0+H_{\rm int}\\
 &=&\sum\limits_i\{\varepsilon (n_{i\uparrow}+n_{i\downarrow})+
 \tilde{\varepsilon}(\tilde{n}_{i\uparrow}+\tilde{n}_{i\downarrow})-
 hS^z_i\}\nonumber \\
 &+&\sum\limits_{ij\sigma}t_{ij}(a^+_{i\sigma}a_{j\sigma}+
 a^+_{i\sigma}\tilde{a}_{j\sigma}+
 \tilde{a}^+_{i\sigma}a_{j\sigma}+\tilde{a}^+_{i\sigma}
 \tilde{a}_{j\sigma}), \nonumber
\end{eqnarray}
where
\begin{equation}
\label{2.3}
\varepsilon =E_0+g/2 ,\qquad
\tilde{\varepsilon} =E_0-g/2
\end{equation}
are energies of the single-site states;
$H_0$ is a single-site (diagonal) term,
$H_{\rm int}$ is a hopping term.

The introduced operators satisfy the following commutation rules:
\begin{eqnarray}
\label{2.4}
\{\tilde{a}^+_{i\sigma},\tilde{a}_{j\sigma'}\}=
\delta_{ij}\delta_{\sigma\sigma'}
\big(\frac{1}{2}-S^z_i\big), \qquad &
\{\tilde{a}^+_{i\sigma},a_{j\sigma'}\}=0, \nonumber\\
\{a^+_{i\sigma},a_{j\sigma'}\}=
\delta_{ij}\delta_{\sigma\sigma'}
\big(\frac{1}{2}+S^z_i\big), \qquad &
\{a^+_{i\sigma},\tilde{a}_{j\sigma'}\}=0.
\end{eqnarray}
In order to calculate the pseudospin mean values we shall use the standard
representation of the statistical operator in the form
\begin{equation}
\label{2.5}
{\rm e}^{-\beta H}={\rm e}^{-\beta H_0}\hat{\sigma}(\beta) ,
\end{equation}
\begin{equation}
\label{2.6}
\hat{\sigma}(\beta)=T_{\tau}\exp\left\{-\int\limits_0^\beta
H_{\rm int}(\tau) {\rm d}\tau \right\},
\end{equation}
which gives the following expressions for $\langle S^z_l\rangle$:
\begin{equation}
\label{2.7}
\langle S^z_l \rangle=\frac{1}{\langle\hat{\sigma}(\beta)\rangle_0}
\langle S^z_l \hat{\sigma}(\beta)\rangle_0=
\langle S^z_l \hat{\sigma}(\beta)\rangle_0^{\rm c} .
\end{equation}
Here, the operators are given in the interaction representation
\begin{equation}
\label{2.8}
A(\tau)={\rm e}^{\tau H_0}A{\rm e}^{-\tau H_0} ,
\end{equation}
the averaging $\langle\dots\rangle_0$ is performed over statistical
distribution with the Hamiltonian $H_0$, and the symbol
$\langle\dots\rangle_0^{\rm c}$
denotes separation of connected diagrams.

\section{Perturbation theory for pseudospin mean values and a diagram
technique}

Expansion of the exponent in $(\ref{2.6})$ in powers of $H_{\rm int}$
$(\ref{2.2})$ leads, after substitution in equation $(\ref{2.7})$, to
an expression that has the form of the sum of infinite series
with terms containing
the averages of the $T$-products of the electron creation
(annihilation) operators $(\ref{2.1})$.
The evaluation of such averages can be made
using the Wick theorem.

In our case this theorem has some differences from the standard formulation.
Namely, each pairing of operators $(\ref{2.1})$ contains
operator factors, i.e.
\begin{eqnarray}
\label{3.3}
&&\stackrel{\line(0,-1){.3}\vector(-1,0){1.7}\line(-1,0){.8}\line(0,-1){.3}\hspace{2em}\null}
{a_i(\tau')a_o^+(\tau)}
=\breve{g}(\tau'-\tau)\delta_{io}P^+_i , \quad
\stackrel{\line(0,-1){.3}\vector(-1,0){1.7}\line(-1,0){.8}\line(0,-1){.3}\hspace{2em}\null}
{\tilde{a}_i(\tau')\tilde{a}_o^+(\tau)}
=\tilde{g}(\tau'-\tau)\delta_{io}P^-_i ,
\\
\nonumber
&&\stackrel{\line(0,-1){.3}\vector(1,0){1.7}\line(1,0){.8}\line(0,-1){.3}\hspace{2em}\null}
{a_o^+(\tau)a_i(\tau')}
=-\breve{g}(\tau'-\tau)\delta_{io}P^+_i,
\quad
\stackrel{\line(0,-1){.3}\vector(1,0){1.7}\line(1,0){.8}\line(0,-1){.3}\hspace{2em}\null}
{\tilde{a}_o^+(\tau)\tilde{a}_i(\tau')}
=-\tilde{g}(\tau'-\tau)\delta_{io}P^-_i.
\end{eqnarray}
Finally, this gives the possibility to express the result in terms of the
products of nonperturbed Green functions
\begin{equation}
\fl
\label{3.4}
\breve{g}_{io}(\tau-\tau')=
\frac{\langle T_\tau a_i(\tau)a_o^+(\tau')\rangle_0}
{\langle \{a_ia_o^+\}\rangle_0}={\rm e}^{\varepsilon(\tau'-\tau)}
\delta_{oi} \left\{
\begin{array}{rl}
(1+{\rm e}^{-\beta\varepsilon})^{-1} & ,\, \tau>\tau'\, ,\\
-(1+{\rm e}^{\beta\varepsilon})^{-1} & , \,\tau'>\tau\, ,
\end{array}
\right.
\end{equation}
$$
\fl
\tilde{g}_{io}(\tau-\tau')=\frac{\langle T_\tau \tilde{a}_i(\tau)
\tilde{a}_o^+(\tau')\rangle_0}{\langle \{\tilde{a}_i
\tilde{a}_o^+\}\rangle_0}=
{\rm e}^{\tilde{\varepsilon}(\tau'-\tau)}
\delta_{oi} \left\{
\begin{array}{rl}
(1+{\rm e}^{-\beta\tilde{\varepsilon}})^{-1} & , \,\tau>\tau'\, ,\\
-(1+{\rm e}^{\beta\tilde{\varepsilon}})^{-1} & , \,\tau'>\tau\, ,
\end{array}
\right.
$$
$$
\fl
\tilde{g}_{io}(\tau-\tau')=\tilde{g}(\tau-\tau')\delta_{io},\qquad
\breve{g}_{io}(\tau-\tau')=\breve{g}(\tau-\tau')\delta_{io},
$$
and averages of a certain number of the projection operators
\begin{equation}
P^+_i=\frac{1}{2}+S^z_i ,\qquad P^-_i=\frac{1}{2}-S^z_i.
\end{equation}
Let us demonstrate this procedure
for one of the terms which appear in the
fourth order of the perturbation theory for $\langle S^z_l\rangle$:
\begin{eqnarray}
\label{3.6}
\lefteqn{
\int\limits_0^\beta \!{\rm d}\tau_1 \int\limits_0^\beta
\!{\rm d}\tau_2
\int\limits_0^\beta \!{\rm d}\tau_3 \int\limits_0^\beta
\!{\rm d}\tau_4
\sum_{iji_1j_1}\sum_{i_2j_2i_3j_3}t_{ij}t_{i_1j_1}t_{i_2j_2}t_{i_3j_3}
}
\\ \nonumber
&&\times\langle T_\tau S^z_l a^+_i(\tau_1) a_j(\tau_1)
\tilde{a}^+_{i_1}(\tau_2)a_{j_1}(\tau_2) a^+_{i_2}(\tau_3)
\tilde{a}_{j_2}(\tau_3)a^+_{i_3}(\tau_4) a_{j_3}(\tau_4) \rangle_0 .
\end{eqnarray}

The stepwise pairing of a certain operator with the other ones gives the
possibility to reduce expression $(\ref{3.6})$ to the sum of the averages of
a smaller number of operators
\begin{eqnarray}
\label{3.7}
\vspace*{1em}
&&\langle T_\tau S^z_l a^+_i(\tau_1) a_j(\tau_1) \tilde{a}^+_{i_1}(\tau_2)
a_{j_1}(\tau_2) a^+_{i_2}(\tau_3) \tilde{a}_{j_2}(\tau_3)
a^+_{i_3}(\tau_4) a_{j_3}(\tau_4) \rangle_0 \nonumber \\
&& \quad=\langle T_{\tau} S^z_l
\stackrel{\line(0,-1){.3}\vector(1,0){11.4}\line(1,0){8.6}\line(0,-1){.3}\hspace{2.3em}\null}
{a^+_i(\tau_1) a_j(\tau_1) \tilde{a}^+_{i_1}(\tau_2)
a_{j_1}(\tau_2) a^+_{i_2}(\tau_3) \tilde{a}_{j_2}(\tau_3)
a^+_{i_3}(\tau_4) a_{j_3}(\tau_4)} \rangle_0 \nonumber \\
&&\qquad+\langle T_\tau S^z_l
\stackrel{\line(0,-1){.3}\vector(1,0){4.9}\line(1,0){3.4}\line(0,-1){.3}\hspace{2.3em}\null}
{a^+_i(\tau_1) a_j(\tau_1) \tilde{a}^+_{i_1}(\tau_2)
a_{j_1}(\tau_2) }
a^+_{i_2}(\tau_3) \tilde{a}_{j_2}(\tau_3)
a^+_{i_3}(\tau_4) a_{j_3}(\tau_4) \rangle_0 \nonumber \\
&&
\vphantom{\Bigg]}
\quad=-\breve{g}_{ij_3}(\tau_1-\tau_4)\langle T_\tau S^z_l P^+_{j_3}
a_j(\tau_1)\tilde{a}^+_{i_1}(\tau_2) a_{j_1}(\tau_2) a^+_{i_2}(\tau_3)
\tilde{a}_{j_2}(\tau_3) a^+_{i_3}(\tau_4) \rangle_0 \\
&&
\vphantom{\Bigg]}
\qquad-\breve{g}_{ij_1}(\tau_1-\tau_2)\langle T_\tau S^z_l P^+_{j_1}
a_j(\tau_1)\tilde{a}^+_{i_1}(\tau_2)  a^+_{i_2}(\tau_3)
\tilde{a}_{j_2}(\tau_3) a^+_{i_3}(\tau_4) a_{j_2}(\tau_3)\rangle_0 .
\nonumber
\end{eqnarray}
The successive application of the pairing procedure for $(\ref{3.7})$
leads, finally, to
\begin{eqnarray}
\label{3.8}
\fl
&&-\breve{g}_{ij_1}(\tau_1{-}\tau_2) \tilde{g}_{i_1j_2}(\tau_2{-}\tau_3)
\breve{g}_{i_3j}(\tau_4{-}\tau_1) \breve{g}_{i_2j_1}(\tau_3{-}\tau_2)
\langle T_\tau S^z_l P^+_j P^+_{j_1} P^-_{j_2} P^+_{j_3}
\rangle_0\ \nonumber \\
\fl
&&-\breve{g}_{ij_3}(\tau_1{-}\tau_4)
\tilde{g}_{i_1j_2}(\tau_2{-}\tau_3)\breve{g}_{i_2j}(\tau_3{-}\tau_1)
\breve{g}_{i_3j_1}(\tau_4{-}\tau_2)\langle T_\tau S^z_l P^+_j P^+_{j_1}
P^-_{j_2} P^+_{j_3}\rangle_0 \\
\fl
&&+\breve{g}_{ij_3}(\tau_1{-}\tau_4)
\tilde{g}_{i_1j_2}(\tau_2{-}\tau_3)\breve{g}_{i_2j_1}(\tau_3{-}\tau_2)
\breve{g}_{i_3j}(\tau_4{-}\tau_1)\langle T_\tau S^z_l P^+_j P^+_{j_1}
P^-_{j_2} P^+_{j_3}\rangle_0.\nonumber
\end{eqnarray}
We introduce the diagrammatic notations
$$
\raisebox{-.16cm}{\epsfysize .5cm\epsfbox{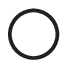}}
-S^z_l;\qquad 1
\raisebox{-.16cm}{\epsfysize .4cm\epsfbox{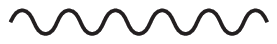}}
1' -t_{11'};
$$
$$
1
\raisebox{-.16cm}{\epsfysize .5cm\epsfbox{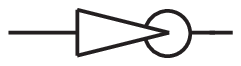}}
1'-(\breve{g}_{11'}P^+_1+\tilde{g}_{11'}P^-_1)
$$
and diagrams
$$
\raisebox{-1.4cm}{\epsfysize 3.cm\epsfbox{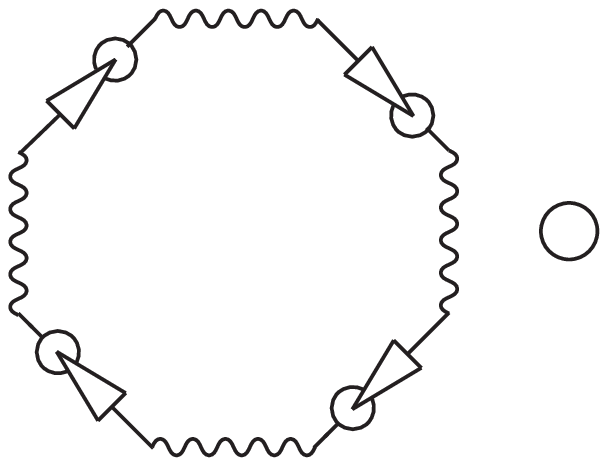}};\qquad
2\times
\raisebox{-1.1cm}{\epsfysize 2.3cm\epsfbox{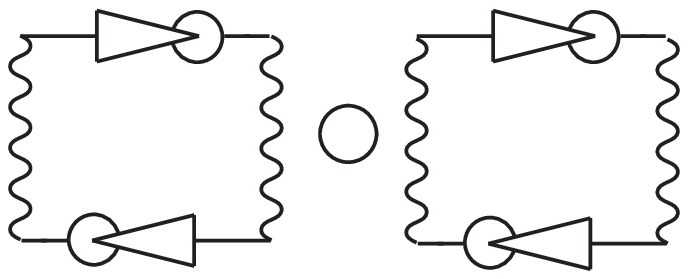}}
$$
which correspond to expression $(\ref{3.8})$.

Expansion of $(\ref{3.8})$ in semi-invariants leads to
multiplication of diagrams (semi-invariants are represented by ovals
surrounding the corresponding vertices with diagonal operators
and contain the $\delta$-symbol on site indexes).
For example,
$$
\epsfxsize 1.\textwidth\epsfbox{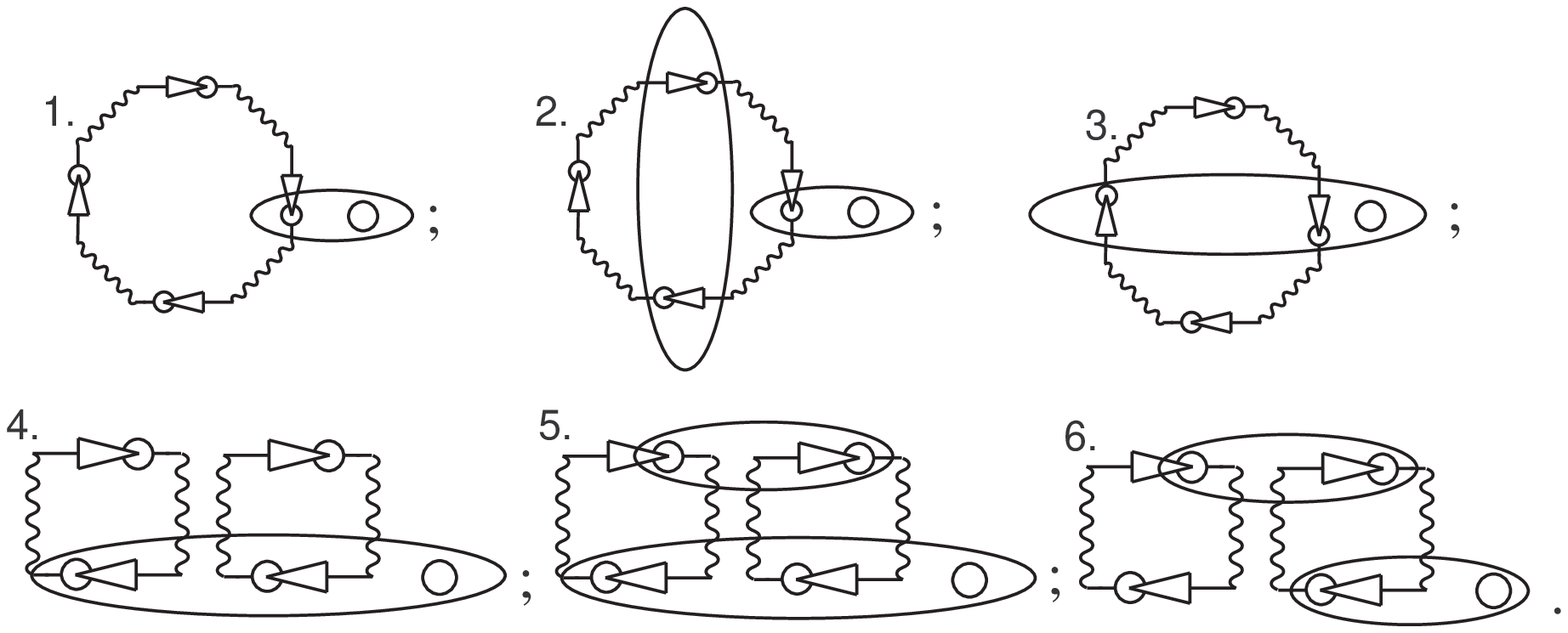}
$$
We shall omit diagrams of types 2, 3, 5, i.e. the types including
semi-in\-va\-ri\-ants of a higher than the first order in the loop (this
means that chain fragments form single-electron Green functions in the
Hubbard-I approximation) and also connection of
two loops by more than one
semi-invariant (this approximation means that a self-consistent
field is taken into account in the zero approximation).

Let us proceed to the momentum-frequency representation in the expressions
for the Green functions determined on a finite interval $0{<}\tau{<}\beta$
when they can be expanded in the Fourier series with discrete frequencies
\begin{equation}
\label{3.9}
\breve{g}(\tau)=\frac{1}{\beta}\sum_n
{\rm e}^{{\rm i}\omega_n \tau}\breve{g}(\omega_n),
\qquad
\tilde{g}(\tau)=\frac{1}{\beta}\sum_n
{\rm e}^{{\rm i}\omega_n \tau}\tilde{g}(\omega_n),
\end{equation}
$$
\breve{g}(\omega_n)=
-\frac{1}{{\rm i}\omega_n-\varepsilon},\hspace*{3em}
\tilde{g}(\omega_n)=
-\frac{1}{{\rm i}\omega_n-\tilde{\varepsilon}},\hspace*{3em}
\omega_n=\frac{2n+1}{\beta}\pi .
$$
The characteristic feature of the already presented diagrams and the diagrams
corresponding to other orders of the perturbation theory is the presence
of chain fragments. The simplest series of chain diagrams is
\begin{equation}
\label{3.10}
\raisebox{-1.6cm}{\epsfysize 2.cm\epsfbox{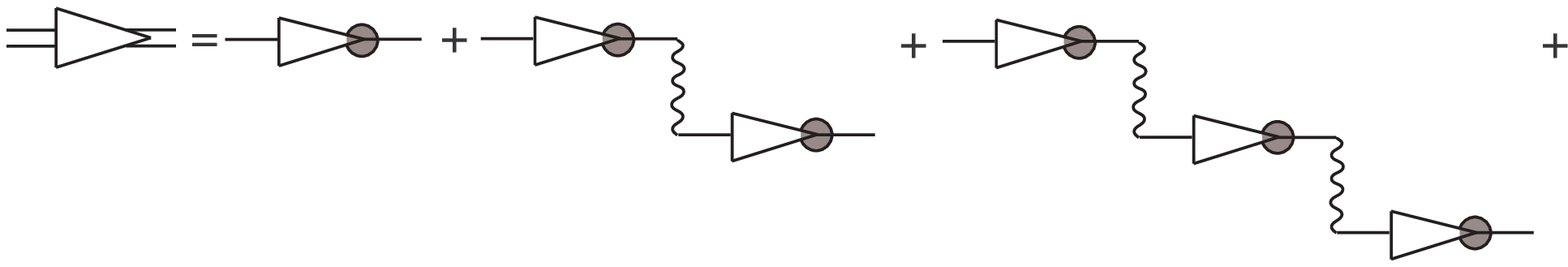}}
\quad
\dots\quad ,
\end{equation}
where
\begin{equation}
\label{3.11}
\raisebox{-0.16cm}{\epsfysize .5cm\epsfbox{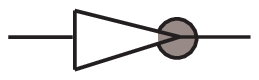}}
=g(\omega_n)=\frac{\langle
P^+\rangle}{{\rm i}\omega_n-\varepsilon}+
\frac{\langle P^-\rangle}{{\rm i}\omega_n-\tilde{\varepsilon}}
\end{equation}
and corresponds to the Hubbard-I approximation for a single-electron
Green function. The expression
\begin{equation}
\label{3.12}
\raisebox{-0.21cm}{\epsfysize .6cm\epsfbox{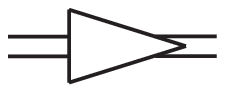}}
=G_{\bk}(\omega_n)=\frac{1}{g^{-1}(\omega_n)-t_{\bk}}
\end{equation}
in the momentum-frequency representation corresponds to the
sum of graphs
$(\ref{3.10})$. The poles of function $G_{\bk}(\omega_n)$ determine the
spectrum of the single-electron excitations
\begin{equation}
\label{3.13}
\varepsilon_{\rs {I},\rs {II}}(t_{\bk})=
\frac{1}{2}(2E_0+t_{\bk})\pm \frac{1}{2}
\sqrt{g^2+4t_{\bk}\langle S^z \rangle g +t^2_{\bk}}\, .
\end{equation}

\begin{figure}[htbp]
\begin{center}
{\epsfysize 7cm\epsfbox{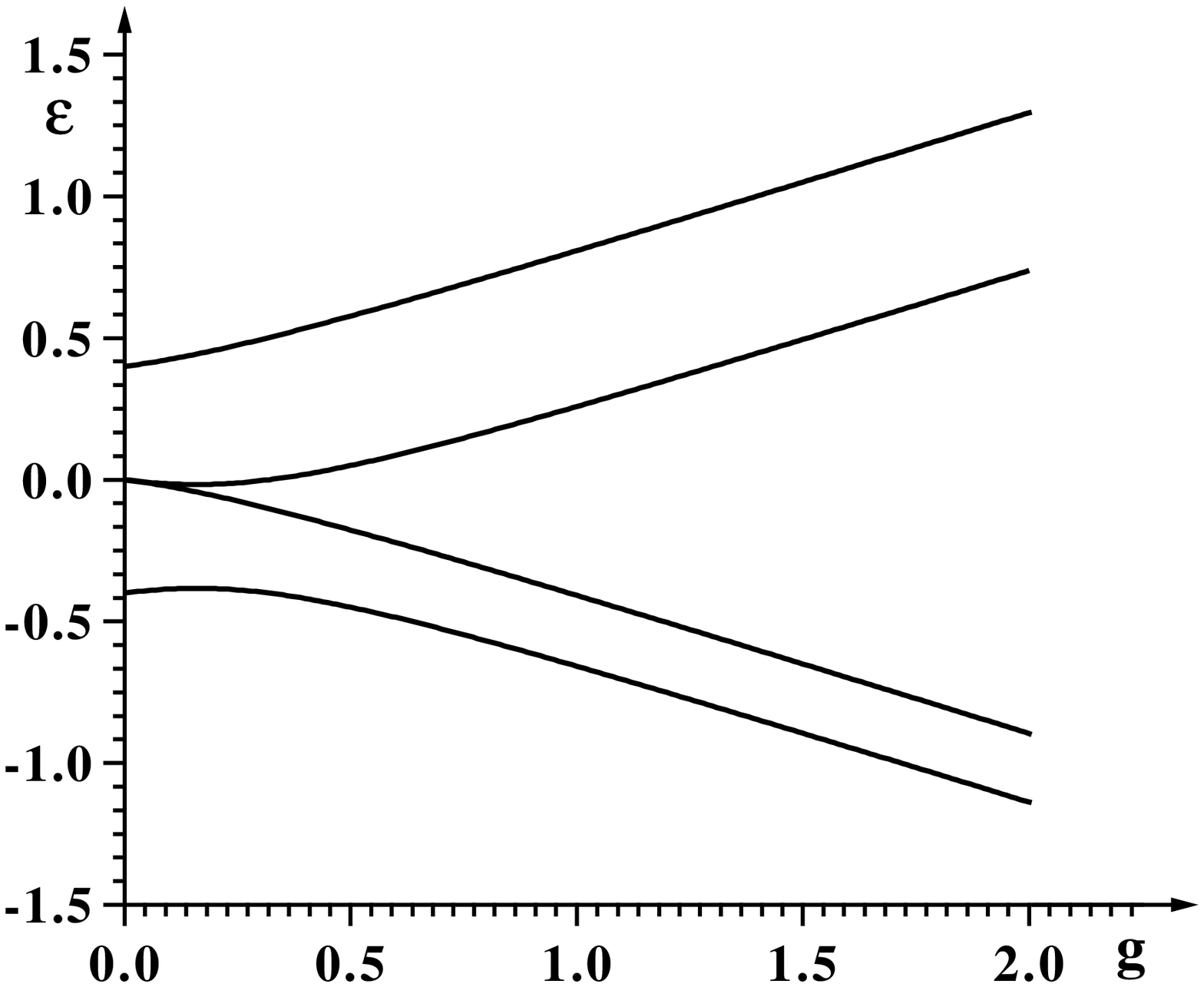}}
\end{center}
\caption{Electron bands boundaries ($W~=~0.4$, $\langle S^z\rangle~=~0.2$).}
\label{1n}
\end{figure}

Behaviour of the electron bands as a function of the coupling constant is
presented in figure~\ref{1n}. One can see that there always
exists a gap in the spectrum. The
widths of subbands depends on the mean value of the pseudospin and in the
case of strong coupling ($g\gg W$) the subbands' halfwidth is equal to
$W\big(\frac{1}{2}\pm \langle S^z\rangle\big) $ ($W$ is the halfwidth of
the initial electron band).

Let us now return to the problem of summation of the diagram series for
the mean value $\langle S^z_l\rangle$
taking into account the above mentioned
arguments. The diagram series has the form
\begin{equation}
\label{3.14}
\langle S^z_l\rangle=
\raisebox{-2.15cm}{\epsfysize 3.5cm\epsfbox{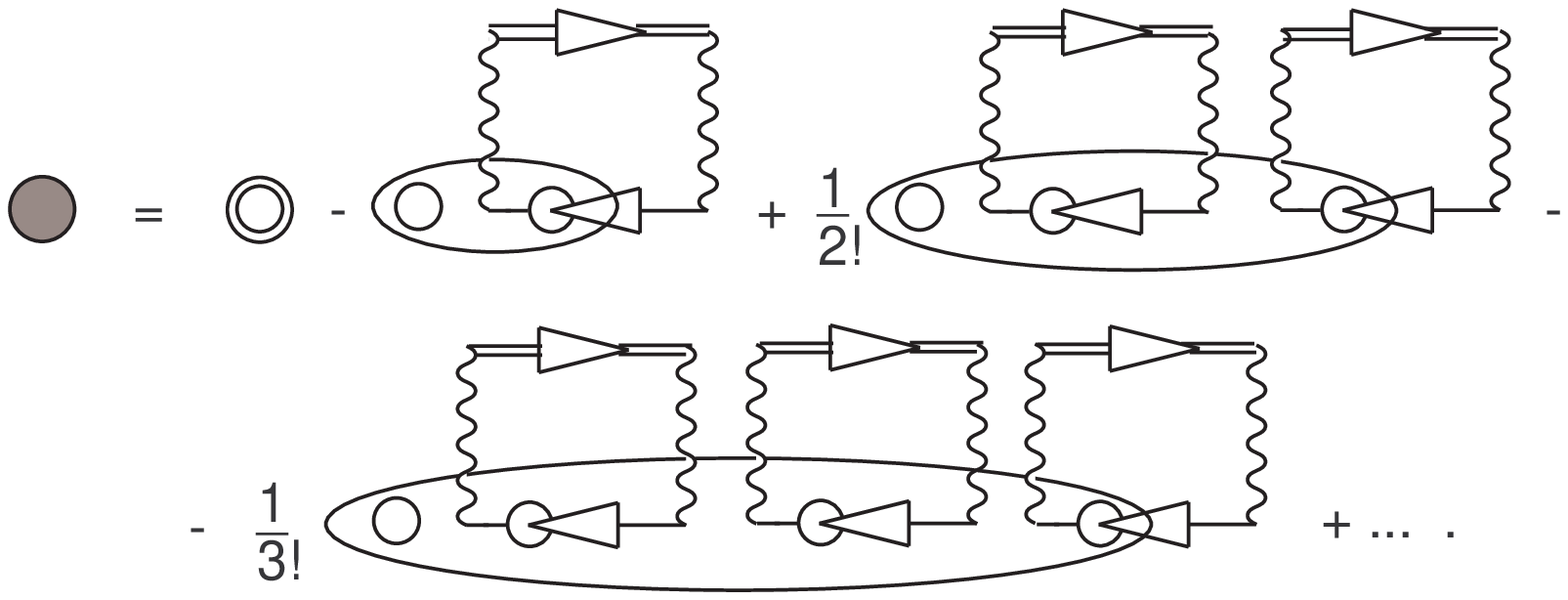}}
\end{equation}
%
The analytical expressions for the loop has the following form
\begin{eqnarray}
\nonumber
\raisebox{-.7cm}{\epsfysize 1.6cm\epsfbox{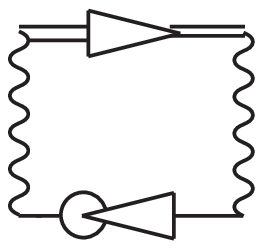}}
&\!=\!&\frac{2}{N}\sum_{n,\bk}\frac{t^2_{\bk}}{g^{-1}(\omega_n)-
t_{\bk}}\left(\frac{P^+_i}{{\rm i}\omega_n-\varepsilon}+
\frac{P^-_i}{{\rm i}\omega_n-\tilde{\varepsilon}}\right)
\\
&\!=\!&\beta(\alpha_1P^+_i+\alpha_2P^-_i) ,
\end{eqnarray}
where we used the notations
$$
\alpha_1=\frac{2}{N\beta}
\sum_{n,\bk}\frac{t^2_{\bk}}{(g^{-1}(\omega_n)-t_{\bk})}
\frac{1}{({\rm i}\omega_n-\varepsilon)} ,
\quad
\alpha_2=\frac{2}{N\beta}
\sum_{n,\bk}\frac{t^2_{\bk}}{(g^{-1}(\omega_n)-t_{\bk})}
\frac{1}{({\rm i}\omega_n-\tilde{\varepsilon})} .
$$
Using decomposition into simple fractions and summation over frequency we
obtain
$$
\alpha_1{=}\frac{2}{N}\sum_{\bk}t_{\bk}
\Big[A_1n(\varepsilon_{\rs I}(t_{\bk}))
+B_1n(\varepsilon_{\rs II}(t_{\bk}))\Big],
\;
\alpha_2{=}\frac{2}{N}\sum_{\bk}t_{\bk}
\Big[A_2n(\varepsilon_{\rs I}(t_{\bk}))
+B_2n(\varepsilon_{\rs II}(t_{\bk}))\Big],
$$
where
$$
A_1=\frac{\varepsilon_{\rs I}(t_{\bk})
-\tilde{\varepsilon}}
{\varepsilon_{\rs I}(t_{\bk})
-\varepsilon_{\rs II}(t_{\bk})} ,
\qquad
B_1=\frac{\varepsilon_{\rs II}(t_{\bk})
-\tilde{\varepsilon}}
{\varepsilon_{\rs II}(t_{\bk})
-\varepsilon_{\rs I}(t_{\bk})} ,
$$
$$
A_2=\frac{\varepsilon_{\rs I}(t_{\bk})
-\varepsilon}
{\varepsilon_{\rs I}(t_{\bk})
-\varepsilon_{\rs II}(t_{\bk})} ,
\qquad
B_2=\frac{\varepsilon_{\rs II}(t_{\bk})
-\varepsilon}
{\varepsilon_{\rs II}(t_{\bk})
-\varepsilon_{\rs I}(t_{\bk})},
$$
and $n(\varepsilon)=\frac{\displaystyle 1}
{\displaystyle 1+{\rm e}^{\beta\varepsilon}}$
is a Fermi distribution.

The equation for $\langle S^z_l\rangle$ can be presented in the form
\begin{eqnarray}
\langle S^z_l\rangle
&\ff=\ff&\langle S^z_l\rangle_0-\langle S^z_l
\beta(\alpha_1P^+_l+\alpha_2P^-_l)\rangle^{\rm c}_0
\nonumber \\
&&+\frac{1}{2!}\langle S^z_l
\beta^2(\alpha_1P^+_l+\alpha_2P^-_l)^2\rangle^{\rm c}_0-\dots
=\langle S^z_l{\rm e}^{-\beta(\alpha_1P^+_l+\alpha_2P^-_l)}\rangle^{\rm c}_0.
\end{eqnarray}
Let us introduce
$$
H_{{\rm MF}}=\sum\limits_iH_i^{{\rm MF}} ,
$$
where
$$
H_i^{{\rm MF}}=H_{i\,0}+\alpha_1P^+_i+\alpha_2P^-_i .
$$
Then, the analytical equation for $\langle S^z_l\rangle$ can be expressed in
the form
\begin{eqnarray}
\label{3.15}
\langle S^z_l\rangle=\langle S^z_l\rangle_{{\rm MF}}
&\!=\!&\frac{{\rm Sp}(S^z_l{\rm e}^{-\beta H_{{\rm MF}}})}
{{\rm Sp}({\rm e}^{-\beta H_{{\rm MF}}})}
\nonumber \\
&\!=\!&\frac{1}{2}\tanh\left\{\frac{\beta}{2}(h+\alpha_2-\alpha_1)+
\ln{\frac{1+{\rm e}^{-\beta\varepsilon}}
{1+{\rm e}^{-\beta\tilde{\varepsilon}}}}
\right\} .
\end{eqnarray}
The difference $\alpha_2-\alpha_1$ corresponds to an internal effective
self-consistent field acting on the pseudospin
\begin{equation}
\label{3.16}
\alpha_2-\alpha_1=\frac{2}{N}\sum_{\bk}t_{\bk}
\frac{\varepsilon-\tilde{\varepsilon}}
{\varepsilon_{\rs I}(t_{\bk})
-\varepsilon_{\rs II}(t_{\bk})}
\Big[n(\varepsilon_{\rs II}(t_{\bk}))
-n(\varepsilon_{\rs I}(t_{\bk}))\Big] .
\end{equation}

\section{The mean value of the particle number}

The diagram series for the mean value $\langle n_i\rangle$
(using the perturbation
theory, the Wick theorem and expansion in semi-invariants) can be presented
in the form
\begin{equation}
\label{4.1}
\langle n_i\rangle
=
\raisebox{-3.cm}{\epsfysize 4.5cm\epsfbox{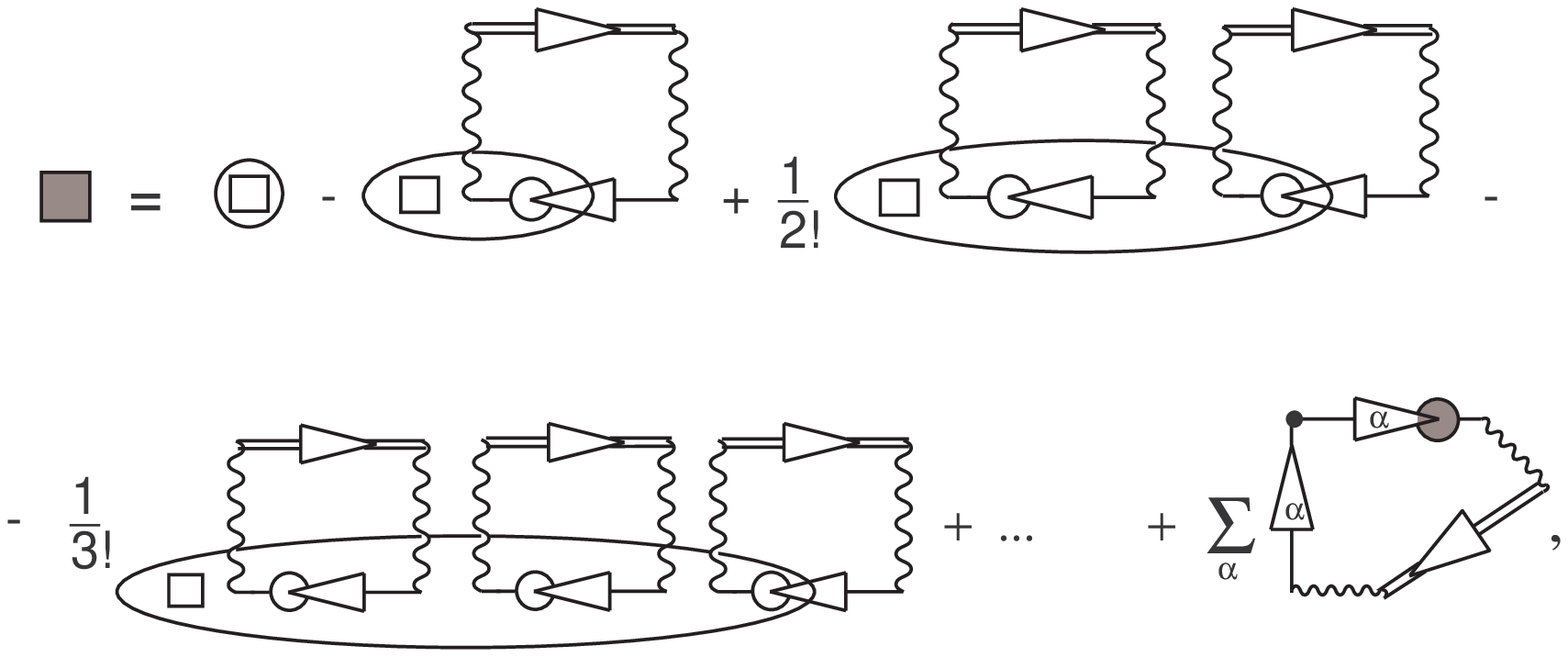}}
\end{equation}
where
$$
\raisebox{-0.16cm}{\epsfysize .5cm\epsfbox{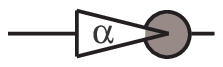}}
=\frac{\langle P^{\alpha}\rangle}{{\rm i}\omega_n-\varepsilon^{\alpha}},
\qquad
\raisebox{-0.16cm}{\epsfysize .5cm\epsfbox{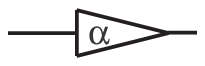}}
=\frac{1}{{\rm i}\omega_n-\varepsilon^{\alpha}},\qquad
\raisebox{-0.16cm}{\epsfysize .5cm\epsfbox{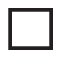}}
=\hat{n}_i,
$$
$$
P^{\alpha}=(P^+;P^-),
\qquad
\varepsilon_{\alpha}=(\varepsilon;\tilde{\varepsilon})
$$
and the last term appears due to the pairing of the electron creation
(annihilation) operators with the operator of the particle number.

An analytical expression for $(\ref{4.1})$ can be obtained starting from
formulae $(\ref{3.11})$, $(\ref{3.12})$:
\begin{equation}
\label{4.2}
\langle n_i \rangle=\langle n_i \rangle_{{\rm MF}}+\frac{2}{N\beta}
\sum_{n,\bk,\alpha}\frac{t^2_{\bk}}{(g^{-1}(\omega_n)-t_{\bk})}
\frac{\langle P^{\alpha}\rangle}{({\rm i}\omega_n-\varepsilon^{\alpha})^2}.
\end{equation}

Using decomposition into simple fractions and summation over frequency
we can present the mean value $\langle n_i \rangle$
in the form:
\begin{equation}
\label{4.4}
\langle n_i \rangle=\frac{2}{N}
\sum_{\bk}\Big[n(\varepsilon_{\rs I}(t_{\bk}))
+n(\varepsilon_{\rs II}(t_{\bk}))\Big]
-2\langle P^+\rangle
n(\tilde{\varepsilon})-2\langle P^-\rangle
n(\varepsilon) .
\end{equation}

\section{Thermodynamic potential}

In order to calculate the thermodynamic potential let us introduce
parameter $\lambda\in [0,1]$ in the initial Hamiltonian
\begin{equation}
\label{5.1}
H_{\lambda}=H_0+\lambda H_{\rm int} ,
\end{equation}
such that $H\to H_0$ for
$\lambda=0$ and $H\to H_0+H_{\rm int}$ for
$\lambda=1$.

Hence,
$$
Z_{\lambda}={\rm Sp}({\rm e}^{-\beta H_{\lambda}})=
{\rm Sp}({\rm e}^{-\beta H_0}\hat{\sigma}_{\lambda}(\beta))=
Z_0\langle\hat{\sigma}_{\lambda}(\beta)\rangle_0 ,
$$
where
$$
\hat{\sigma}_{\lambda}(\beta)=
T_{\tau}\exp\left\{\!-\lambda\int\limits_0^\beta
\! H_{\rm int}(\tau) {\rm d}\tau \right\} ,
$$
and
\begin{equation}
\label{5.2}
\Omega_{\lambda}=-\frac{1}{\beta}\ln Z_0
-\frac{1}{\beta}\ln\langle\hat{\sigma}_{\lambda}(\beta)\rangle_0 ,
\end{equation}
$$
\Delta\Omega_{\lambda}=\Omega_{\lambda}-\Omega_0=
-\frac{1}{\beta}\ln\langle\hat{\sigma}_{\lambda}(\beta)\rangle_0 .
$$
Here $\Omega_0$ is a thermodynamic potential calculated with the
single-site (diagonal) part of the initial Hamiltonian.

Therefore,
\begin{equation}
\label{5.3}
\Delta\Omega=\int\limits_0^1 {\rm d}\lambda
\! \left(\! \frac{{\rm d}\Omega_{\lambda}}{{\rm d}\lambda}\! \right) .
\end{equation}
For value ${\rm d}\Omega_{\lambda}/{\rm d}\lambda$, we can immediately
write the diagram series in the next form:
\begin{equation}
\label{5.4}
\beta \frac{{\rm d}\Omega_{\lambda}}{{\rm d}\lambda}=
\raisebox{-1.2cm}{\epsfysize 2.6cm\epsfbox{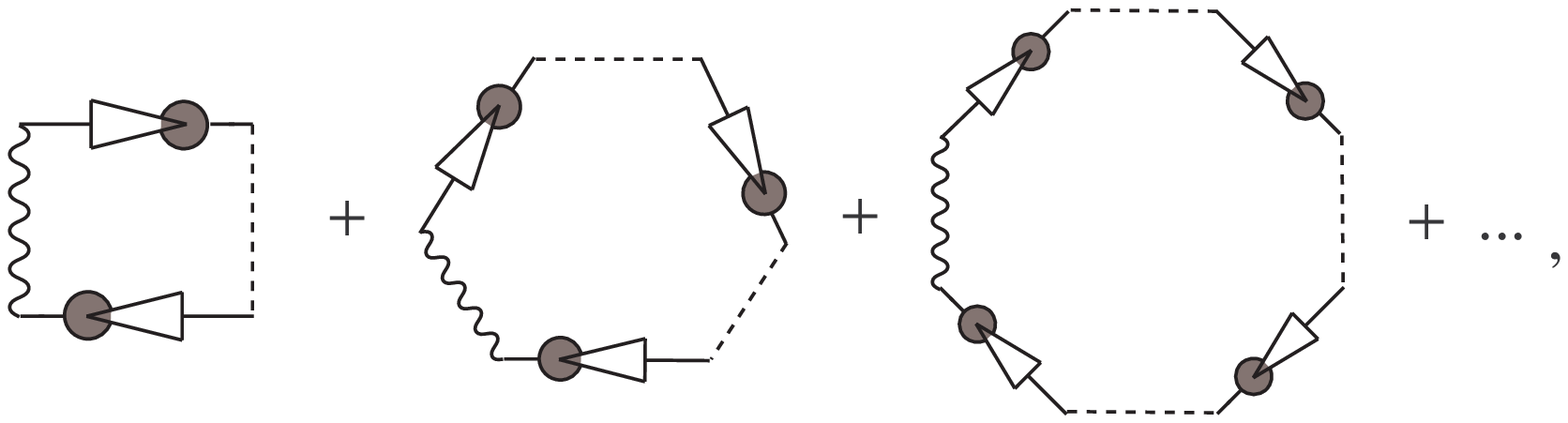}}
\end{equation}
where
\raisebox{-0.6cm}{\epsfysize 1.4cm\epsfbox{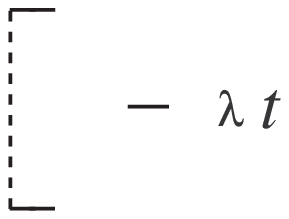}},
and also
$$
{\epsfysize 4.5cm\epsfbox{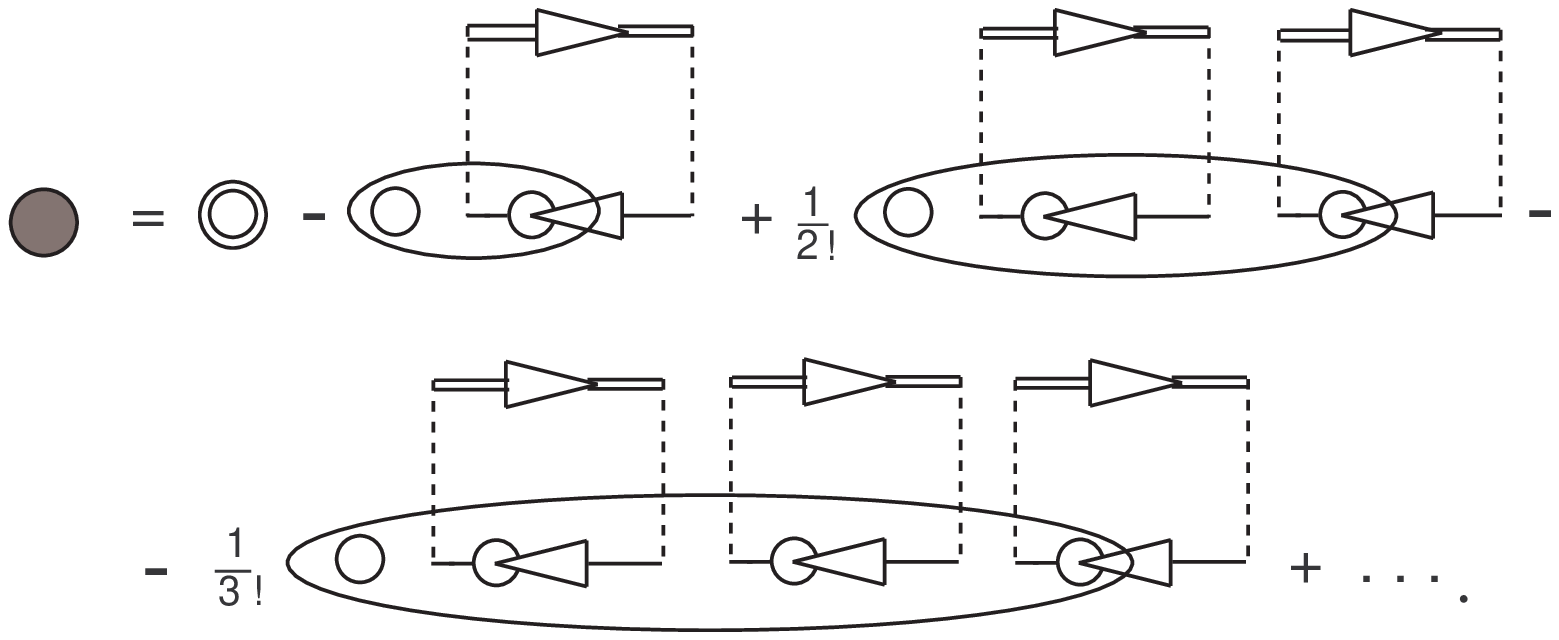}}
$$

Expression $(\ref{5.3})$ can be presented in the form (using the diagram
series $(\ref{5.4})$):
\begin{eqnarray}
\Delta\Omega
&\!=\!&\frac{2}{N\beta}\sum_{n,\bk}
\int\limits_0^1
\lambda t^2_{\bk}g^2_{\lambda}(\omega_n)
\frac{1}{1-\lambda t_{\bk}g_{\lambda}(\omega_n)}
{\rm d}\lambda \nonumber \\
\label{5.5}
&\!=\!&-\frac{2}{N\beta}\sum_{n,\bk}\ln(1-t_{\bk}g(\omega_n))-
\frac{2}{N\beta}\sum_{n,\bk}\int\limits_0^1
\frac{\lambda t_{\bk}\frac{{\rm d}g_{\lambda}(\omega_n)}{{\rm d}\lambda}}
{1-\lambda t_{\bk}g_{\lambda}(\omega_n)} {\rm d}\lambda .
\end{eqnarray}
The first term in expression $(\ref{5.5})$ may be written in a diagram
form as
\begin{equation}
\label{5.6}
{\epsfysize 2.6cm\epsfbox{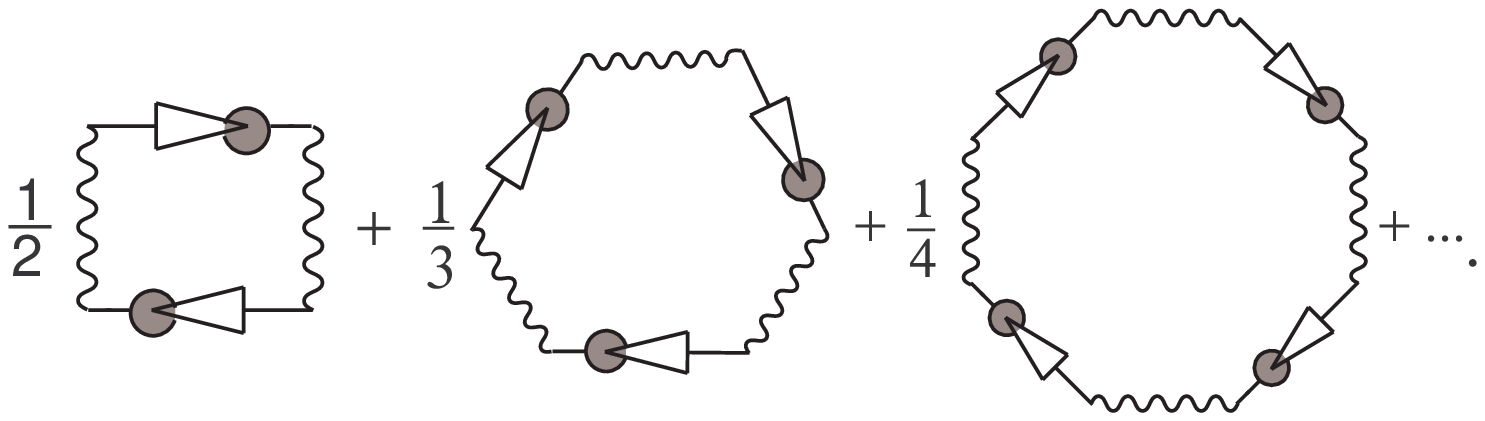}}
\end{equation}
Series $(\ref{5.6})$ describes an electron gas whose
energy spectrum is
defined by the total pseudospin field. This series is in conformity with
the so-called one-loop approximation.

The second term in expression $(\ref{5.5})$ can be integrated to the
following diagram series
\begin{equation}
\label{5.7}
{\epsfysize 3.6cm\epsfbox{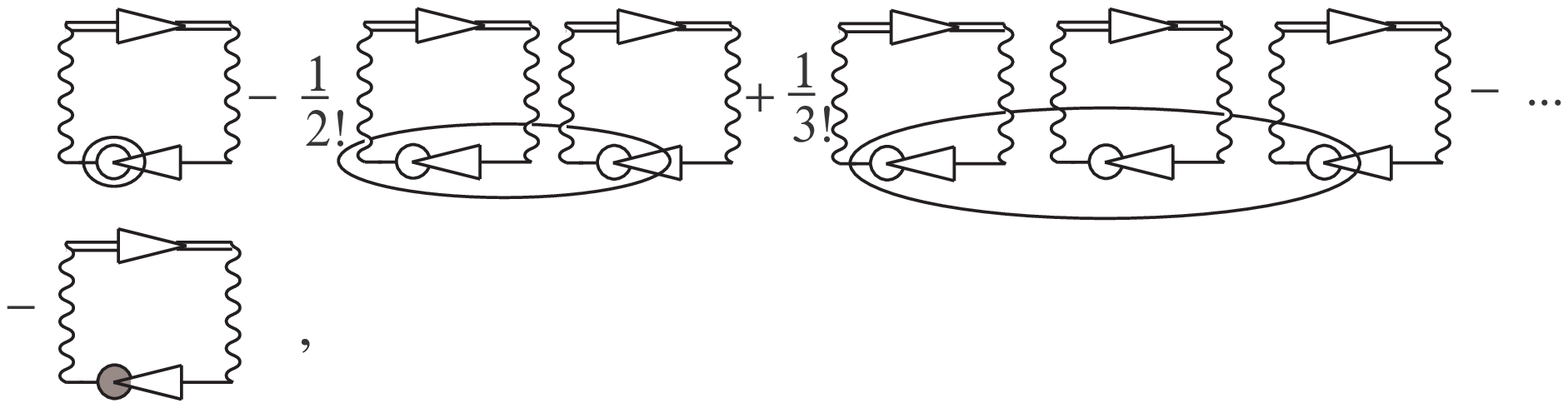}}
\end{equation}
and appears due to the presence of a pseudospin subsystem.

Finally, the diagram series for $\beta\Delta\Omega $ can be written as
a sum of expressions $(\ref{5.6})$ and $(\ref{5.7})$, and
the corresponding analytical expression is the following:
\begin{eqnarray}
\label{5.9}
\Delta\Omega
&\!\!=\!\!&-\frac{2}{N\beta}\sum_{\bk}
\ln\frac{(\cosh\frac{\beta}{2}\varepsilon_{\rs I}(t_{\bk}))
(\cosh\frac{\beta}{2}\varepsilon_{\rs II}(t_{\bk}))}
{(\cosh\frac{\beta}{2}\varepsilon)
(\cosh\frac{\beta}{2}\tilde{\varepsilon})}
\nonumber \\
&&-\frac{1}{\beta}\ln \cosh\left\{
\frac{\beta}{2}(h+\alpha_2-\alpha_1)+\ln\frac
{1+{\rm e}^{-\beta\varepsilon}}
{1+{\rm e}^{-\beta\tilde{\varepsilon}}}
\right\}
\nonumber \\
&&+\frac{1}{\beta}\ln \cosh\left\{
\frac{\beta}{2} h+\ln\frac
{1+{\rm e}^{-\beta\varepsilon}}
{1+{\rm e}^{-\beta\tilde{\varepsilon}}}
\right\}+\langle S^z\rangle(\alpha_2-\alpha_1) .
\end{eqnarray}
Here, decomposition in simple fractions and summation over
frequency were done.

Then, since the thermodynamic potential is a function of the argument
$\langle S^z\rangle$, let us check the consistency of approximations made
for $\langle S^z\rangle$, $\langle n\rangle$ and thermodynamic
potential $\Omega $.
For this purpose let us derive the mean values
$\langle S^z\rangle$ and
$\langle n\rangle$ from the expression
for the grand thermodynamic potential
$$
\frac{{\rm d}\Omega}{{\rm d}(-\mu)}=\frac{2}{N}
\sum_{\bk}\Big[n(\varepsilon_{\rs I}(t_{\bk}))
+n(\varepsilon_{\rs II}(t_{\bk}))\Big]-2\langle P^+\rangle
n(\tilde{\varepsilon})-2\langle P^-\rangle
n(\varepsilon) ,
$$
$$
\frac{{\rm d}\Omega}{{\rm d}(-h)}=
\frac{1}{2}\tanh\left\{\frac{\beta}{2}(h+\alpha_2-\alpha_1)+
\ln{\frac{1+{\rm e}^{-\beta\varepsilon}}
{1+{\rm e}^{-\beta\tilde{\varepsilon}}}}
\right\} .
$$
We thus obtain
$$
\frac{{\rm d}\Omega}{{\rm d}(-\mu)}=\langle n \rangle ,\hspace{4em}
\frac{{\rm d}\Omega}{{\rm d}(-h)}=\langle S^z \rangle .
$$
Therefore, the calculation of the mean values of the pseudospin and particle
number operators as well as the thermodynamic potential is performed in
the same approximation which corresponds to the mean field one.

\section{Pseudospin, electron and mixed correlators}

In this section our aim is to calculate the correlators
\begin{eqnarray*}
K^{ss}_{lm}(\tau-\tau')&=&
\langle T\tilde{S}^z_l(\tau)\tilde{S}^z_m(\tau')
\rangle^{\rm c} ,\\
K^{sn}_{lm}(\tau-\tau')&=&
\langle T\tilde{S}^z_l(\tau)\tilde{n}_m(\tau')
\rangle^{\rm c} ,\\
K^{nn}_{lm}(\tau-\tau')&=&
\langle T\tilde{n}_l(\tau)\tilde{n}_m(\tau')
\rangle^{\rm c} ,\\
\end{eqnarray*}
constructed of the operators given in the Heisenberg representation with
an imaginary time argument.

Let us present a diagram series for the correlation function (in the
momentum-frequency representation) within a self-consistent scheme
in the framework of the generalized random phase approximation (GRPA)
(which was applied in \cite{9,10} where the magnetic susceptibility of the
ordinary Hubbard model and $t-J$ model was considered).
In our case (the absence of the Hubbard correlation)
this approximation is reduced, because the so-called ladder diagrams
\cite{11} with antiparallel lines disappear.

We would like to remind that we have omitted diagrams including
semi-in\-va\-ri\-ants of a higher than the first order in the loop and
also connection of two loops by more than one semi-invariant.
\begin{equation}
\label{6.1}
\langle S^zS^z\rangle_{\bq}=
\raisebox{-1.1cm}{\epsfysize 2.3cm\epsfbox{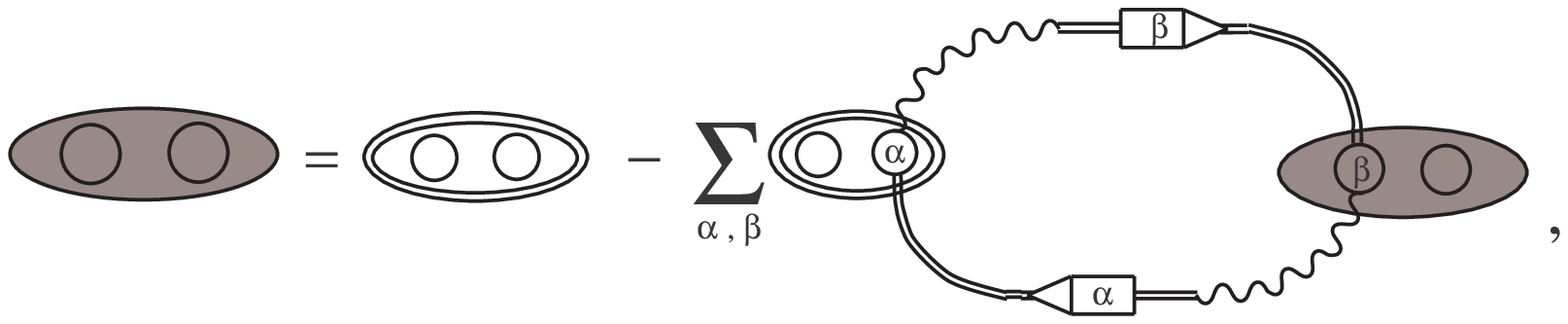}}
\end{equation}
where we define
$$
\raisebox{-1.4cm}{\epsfysize 1.7cm\epsfbox{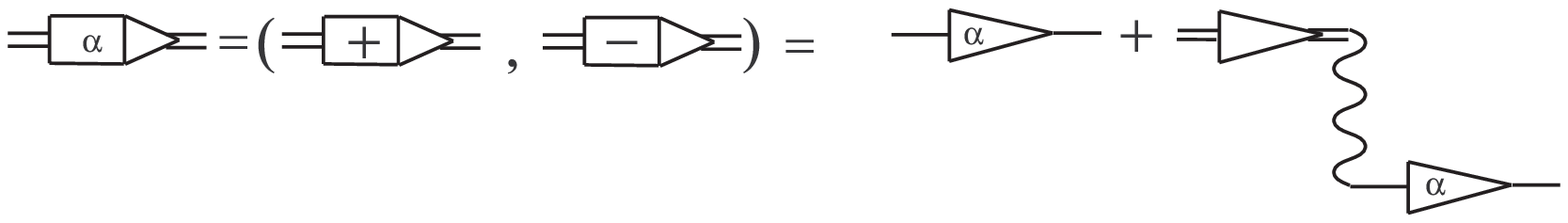}}
=\Gamma^{\alpha}({\bk},\omega_n);
$$
$$
P^{\alpha}=(P^+,P^-);\quad\alpha=(0,1);\quad\varepsilon^{\alpha}
=(\varepsilon,\tilde{\varepsilon});\quad
\raisebox{-0.2cm}{\epsfysize .6cm\epsfbox{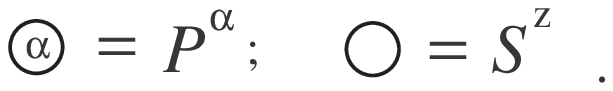}}
$$
Here, the first term in equation $(\ref{6.1})$ takes into account a direct
influence of the internal effective self-consistent field on pseudospins
and is given by
\begin{equation}
\label{6.2}
{\epsfysize 3.8cm\epsfbox{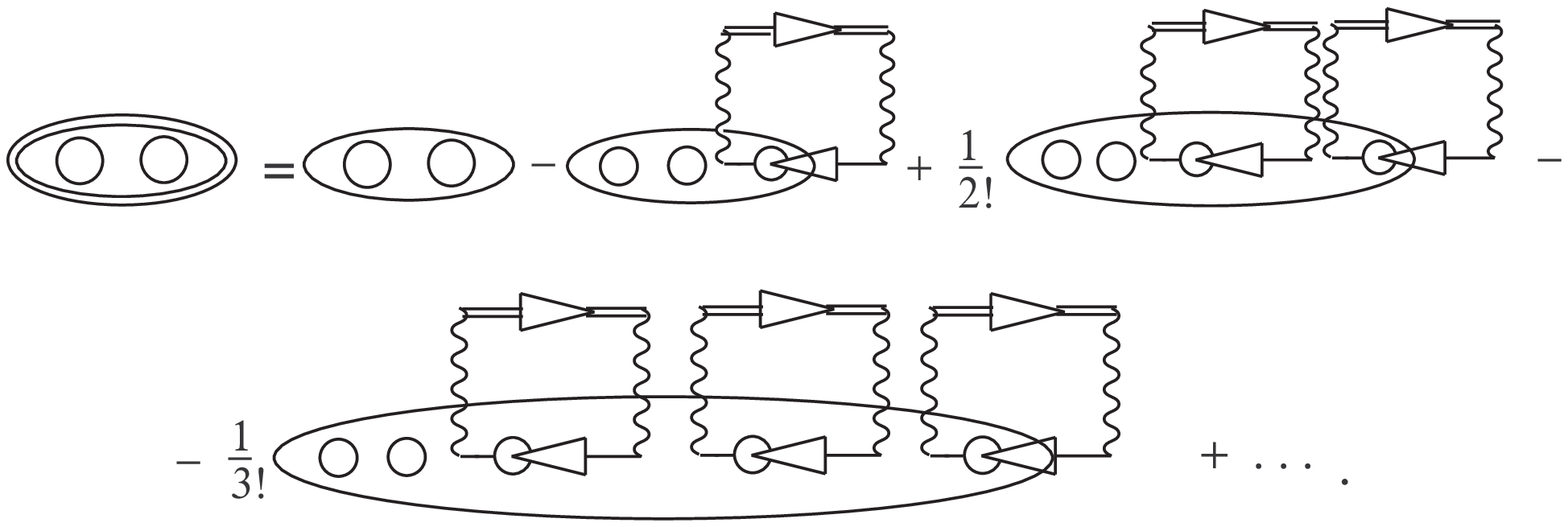}}
\end{equation}
Series $(\ref{6.2})$ means a se\-cond-or\-der se\-mi-in\-va\-ri\-ant
renormalized due to the ``sin\-gle-tail'' parts, and is thus calculated
by $H_{\rm MF}$.

The second term in equation $(\ref{6.1})$ describes an
interaction between
pseudospins which is mediated by electrons (the energy of the electron
spectrum is defined by the total pseudospin field).

We introduce the shortened notations
\begin{equation}
\label{6.3}
\mbox{\fbox{$\Pi$}}^{\alpha,\beta}_{\bq}=
\raisebox{-1.1cm}{\epsfysize 2.3cm\epsfbox{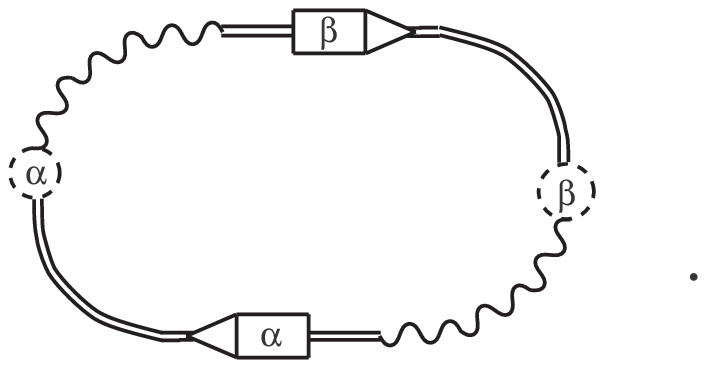}}
\end{equation}
Solution of equation $(\ref{6.1})$ can be written in the analytical form
\begin{equation}
\label{6.4}
\langle S^z S^z\rangle_{\bq}=\frac{1/4-\langle S^z\rangle^2}
{1+\sum\limits_{\alpha,\beta}(-1)^{\alpha+\beta}
\mbox{\fbox{$\Pi$}}^{\alpha,\beta}_{\bq}(\frac{1}{4}-\langle
S^z\rangle^2)} , \end{equation}
where
\begin{equation}
\label{6.5}
\mbox{\fbox{$\Pi$}}^{\alpha,\beta}_{\bq}=
\frac{2}{N}\sum_{n,\bk}t_{\bk}
t_{{\bk}+{\bq}}\Gamma^{\alpha}({\bk},\omega_n)
\Gamma^{\beta}({\bk}+{\bq},\omega_n) ,
\end{equation}
\begin{equation}
\label{6.6}
\Gamma^{\alpha}({\bk},\omega_n)=
\frac{1}{({\rm i}\omega_n-\varepsilon^{\alpha})}
\frac{1}{(1-t_{\bk}g(\omega_n))} .
\end{equation}
Decomposition of function $\Gamma^{\alpha}({\bk},\omega_n)$ into
simple fractions and the subsequent evaluation of the sum over frequency
leads to the next expression:
\begin{eqnarray}
\fll
\sum_{\alpha,\beta}(-1)^{\alpha+\beta}
\mbox{\fbox{$\Pi$}}^{\alpha,\beta}_{\bq}
&\ff=\ff&\frac{2\beta}{N}\sum_{\bk} \frac
{t_{\bk}t_{\bk+{\bq}}
(\varepsilon-\tilde{\varepsilon})^2}
{[\varepsilon_{\rs I}(t_{\bk})
-\varepsilon_{\rs II}(t_{\bk})]
[\varepsilon_{\rs I}(t_{\bk+{\bq}})
-\varepsilon_{\rs II}
(t_{\bk+{\bq}})]}
\nonumber \\
\label{6.7}
&&\times\left\{\frac{n[\varepsilon_{\rs I}(t_{\bk})]-
n[\varepsilon_{\rs I}(t_{\bk+{\bq}})]}
{\varepsilon_{\rs I}(t_{\bk})
-\varepsilon_{\rs I}(t_{\bk+\bq})}+
\frac{n[\varepsilon_{\rs II}(t_{\bk})]-
n[\varepsilon_{\rs II}(t_{\bk+\bq})]}
{\varepsilon_{\rs II}(t_{\bk})-
\varepsilon_{\rs II}(t_{\bk+\bq})}
\right. \nonumber \\
&&\left.
-\frac{n[\varepsilon_{\rs I}(t_{\bk})]
-n[\varepsilon_{\rs II}
(t_{\bk+\bq})]} {\varepsilon_{\rs I}(t_{\bk})
-\varepsilon_{\rs II} (t_{\bk+\bq})}
-\frac{n[\varepsilon_{\rs II}(t_{\bk})]-
n[\varepsilon_{\rs I}(t_{\bk+\bq})]}
{\varepsilon_{\rs II}(t_{\bk})
-\varepsilon_{\rs I}(t_{\bk+\bq})}
\right\} .
\end{eqnarray}
After substitution (\ref{6.7}) in equation (\ref{6.4}) we finally obtain
an expression for\linebreak $\langle S^z S^z \rangle_{\bq}$.

This formula for the uniform case $({\bq}=0)$ can be rewritten
as
\begin{eqnarray}
\label{6.8}
\langle S^zS^z\rangle_{{\bq}=0}
&&\hspace*{-1.6em}=(1/4-\langle S^z\rangle^2)
\\
&&\hspace*{-3.em}\times
\left\{1-\left(
\frac{4\beta}{N}\sum_{\bk}t_{\bk}^2
\frac{(\varepsilon-\tilde{\varepsilon})^2}
{[\varepsilon_{\rs I}(t_{\bk})
-\varepsilon_{\rs II}(t_{\bk})]^3}
\{n[\varepsilon_{\rs I}(t_{\bk})]
-n[\varepsilon_{\rs II}(t_{\bk})]\}
\right.
\right.
\nonumber \\
&&\hspace*{-3.em}+
\left.
\left.
\frac{\beta^2}{2N}\sum_{\bk}
\frac{t_{\bk}^{2} (\varepsilon-\tilde{\varepsilon})^{2}}
{[\varepsilon_{\rs I}(t_{\bk})
-\varepsilon_{\rs II}(t_{\bk})]^{2}}
\left\{\frac{1}{\cosh^{2}\frac{\beta\varepsilon_{\rs I}
(t_{\bk})}{2}}+
\frac{1}{\cosh^{2}\frac{\beta\varepsilon_{\rs II}
(t_{\bk})}{2}}
\right\}\right)(\frac{1}{4}-\langle S^z\rangle^{2})
\right\}^{-1} \!\!\!\!.
\nonumber
\end{eqnarray}
Expression $(\ref{6.8})$ can be obtained from the derivative ${\rm d}\langle
S^z\rangle/ {\rm d}(\beta h)$.
This means that the mean values of the
pseudospin and pseudospin correlators are
derived in the same approximation.

For a mixed correlator the diagram series has the form
\begin{equation}
\label{6.9}
\langle S^zn\rangle_{\bq}=
\raisebox{-.7cm}{\epsfysize 1.7cm\epsfbox{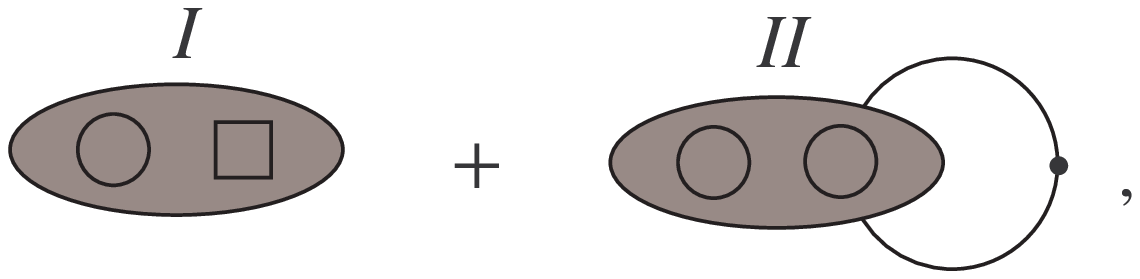}}
\end{equation}
where
\begin{equation}
\label{6.10}
I=
\raisebox{-1.1cm}{\epsfysize 2.3cm\epsfbox{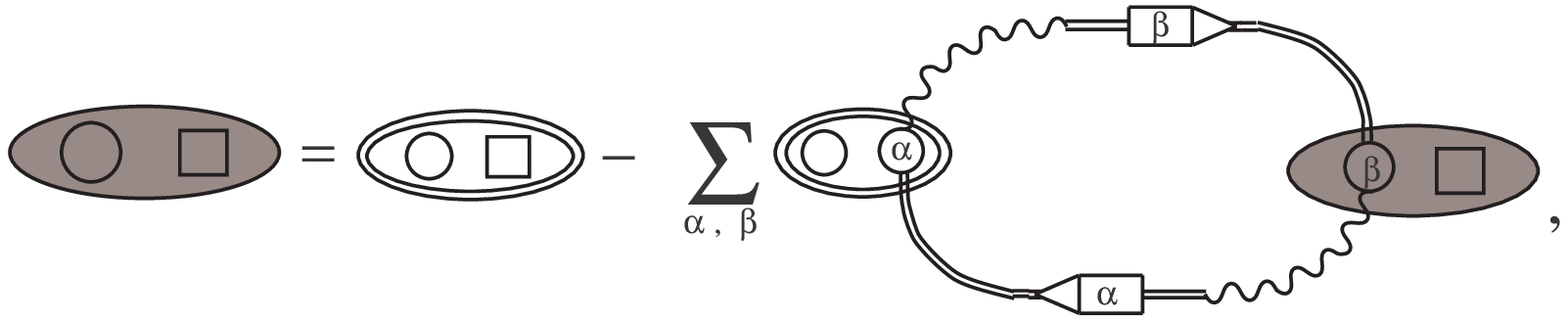}}
\end{equation}
\begin{equation}
\label{6.11}
II=
\raisebox{-1.cm}{\epsfysize 2.1cm\epsfbox{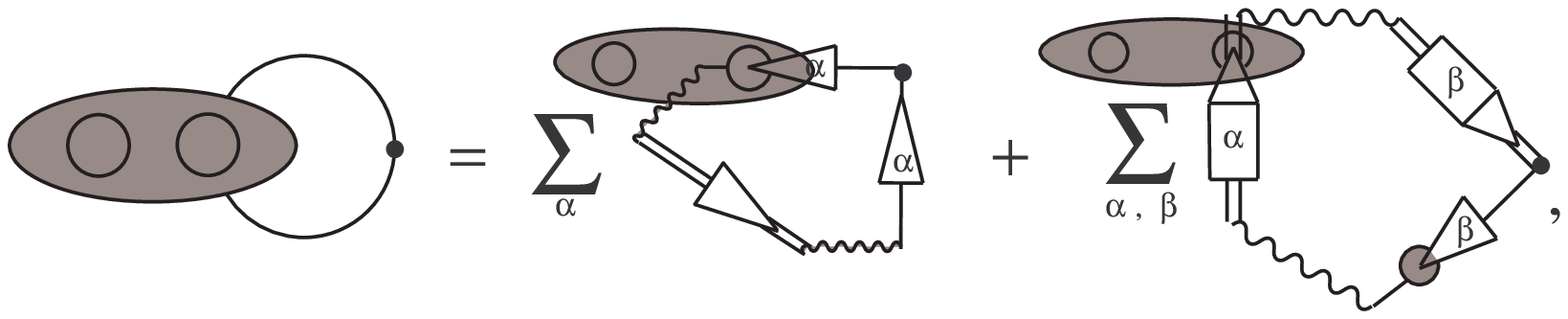}}
\end{equation}
$$
\raisebox{-1.2cm}{\epsfysize 1.5cm\epsfbox{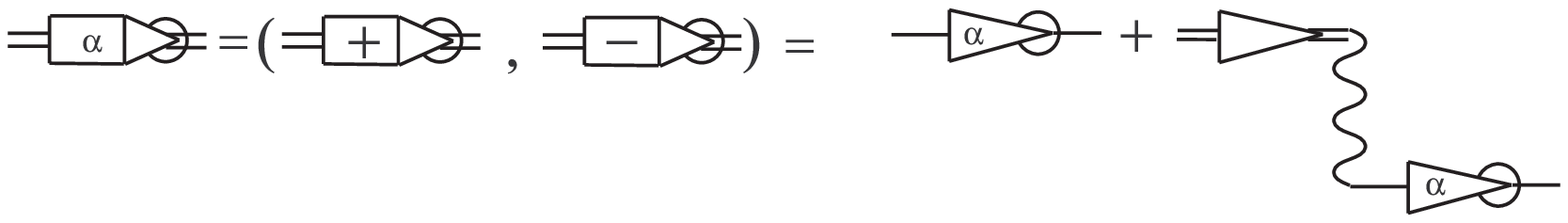}}=
P^{\alpha}\Gamma^{\alpha}({\bk},\omega_n) .
$$
Solution of equation $(\ref{6.10})$ can be written in the analytical form
\begin{equation}
\label{6.12}
I=
2(n(\varepsilon)-n(\tilde{\varepsilon}))\langle S^z S^z\rangle_{\bq} .
\end{equation}
Here we start from formula $(\ref{6.4})$ and from the next relation:
\begin{equation}
\label{6.13}
\frac{\langle S^zn\rangle_{\rm MF}-\langle S^z\rangle\langle n\rangle}
{\frac{1}{4}-\langle S^z \rangle^2}=
2(n(\varepsilon)-n(\tilde{\varepsilon})) .
\end{equation}
The second term in diagram series $(\ref{6.9})$ can be presented as
\begin{eqnarray}
II
&\ff=\ff&
\frac{2}{N}\langle S^z S^z\rangle_{\bq}\sum_{\bk}\frac
{t_{\bk}(\varepsilon-\tilde{\varepsilon})}
{\varepsilon_{\rs I}(t_{\bk})
-\varepsilon_{\rs II}(t_{\bk})}
\nonumber \\
\label{6.14}
&&\times\left[\frac{n[\varepsilon_{\rs I}(t_{\bk})]-
n[\varepsilon_{\rs I}(t_{\bk+\bq})]}
{\varepsilon_{\rs I}(t_{\bk})
-\varepsilon_{\rs I}(t_{\bk+\bq})}+
\frac{n[\varepsilon_{\rs I}(t_{\bk})]-
n[\varepsilon_{\rs II}(t_{\bk+\bq})]}
{\varepsilon_{\rs I}(t_{\bk})
-\varepsilon_{\rs II}(t_{\bk+\bq})}
\right. \nonumber \\
&&\left.
-\frac{n[\varepsilon_{\rs II}(t_{\bk})]-
n[\varepsilon_{\rs I}(t_{\bk+\bq})]}
{\varepsilon_{\rs II}(t_{\bk})
-\varepsilon_{\rs I}(t_{\bk+\bq})}-
\frac{n[\varepsilon_{\rs II}(t_{\bk})]-
n[\varepsilon_{\rs II}(t_{\bk+\bq})]}
{\varepsilon_{\rs II}(t_{\bk})
-\varepsilon_{\rs II}(t_{\bk+\bq})}
\right] .
\end{eqnarray}
Let us introduce the shortened notations for expression $(\ref{6.14})$
\begin{equation}
\label{6.15}
II=\langle S^zS^z\rangle_{\bq} [\oplus]_{\bq} .
\end{equation}
In this way we obtain
\begin{equation}
\label{6.16}
\langle S^z n\rangle_{\bq}=
2\Big[n(\varepsilon)-n(\tilde{\varepsilon})\Big]
\langle S^z S^z\rangle_{\bq}+
\langle S^z S^z\rangle_{\bq} [\oplus]_{\bq} .
\end{equation}
From our diagram series we can see: correlators containing the pseudospin
variable $S^z$ are different from zero only in a static case. This is due
to the fact that operator $S^z$ commutes with the Hamiltonian, being an
integral of motion.

For an electron correlator our diagram series has the form:
\begin{equation}
\label{6.17}
\langle nn\rangle_{\bq,\omega}=
\raisebox{-2.4cm}{\epsfysize 3.2cm\epsfbox{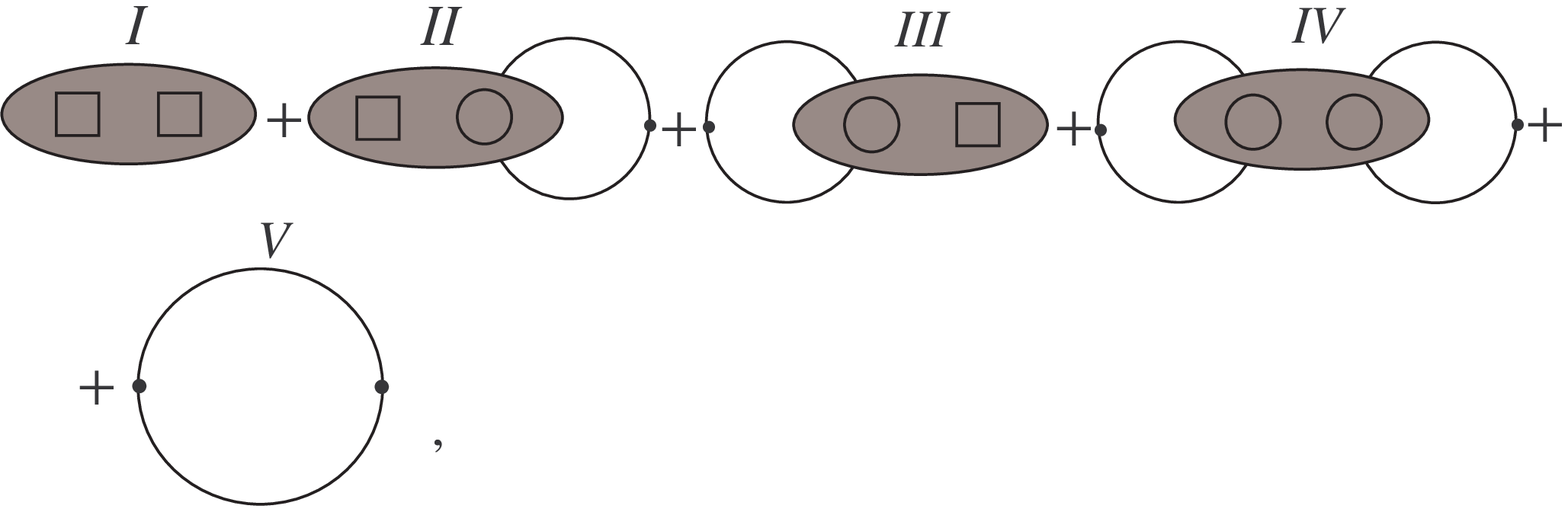}}
\end{equation}
and only the last term is not equal to zero for non-zero frequencies.
Let us consider series $(\ref{6.17})$ term by term.

The first term in series $(\ref{6.17})$ may be written as
\begin{equation}
\label{6.18}
I=
\raisebox{-1.1cm}{\epsfysize 2.3cm\epsfbox{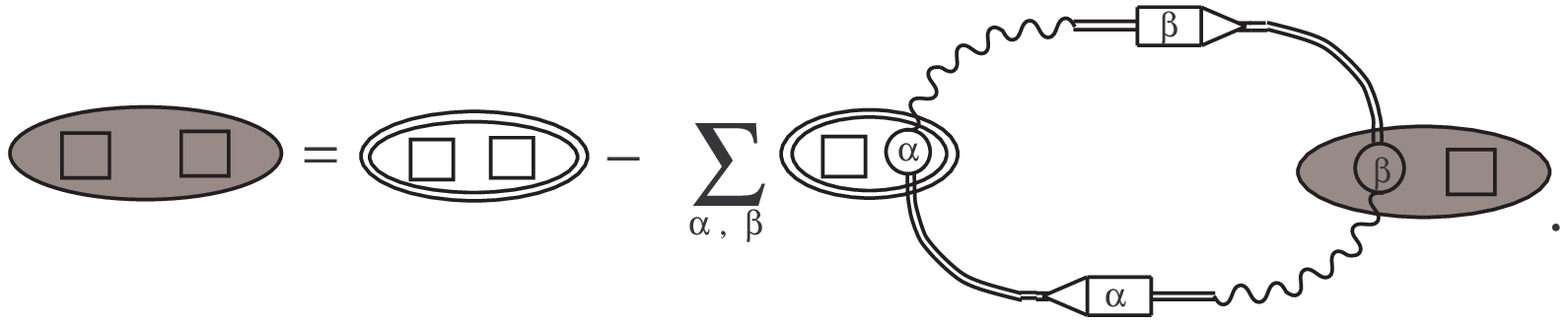}}
\end{equation}
After simple transformation we can obtain the next relation:
\begin{equation}
\label{6.19}
\langle nn \rangle_{\rm MF}
-\langle n \rangle^2-
\frac{1}{2}\left(
\frac{\langle P^+\rangle}{\cosh^2\frac{\beta\varepsilon}{2}}+
\frac{\langle P^-\rangle}{\cosh^2\frac{\beta\tilde{\varepsilon}}{2}}
\right)
=\frac{(
\langle n S^z \rangle_{\rm MF}-\langle n \rangle\langle S^z \rangle)^2}
{\langle P^+ \rangle\langle P^-\rangle} .
\end{equation}
This relation makes it possible to write immediately a simple analytical
expression for series $(\ref{6.18})$
\begin{equation}
\label{6.20} I=\Bigg\{
[2(n(\varepsilon)-n(\tilde{\varepsilon}))]^2
\langle S^zS^z\rangle_{\bq}+\frac{1}{2}\left(
\frac{\langle P^+\rangle}{\cosh^2\frac{\beta\varepsilon}{2}}+
\frac{\langle P^-\rangle}{\cosh^2\frac{\beta\tilde{\varepsilon}}{2}}
\right)\Bigg\}\delta(\omega) .
\end{equation}
Analytical expressions for the $II$-term can be obtained starting from
formulae $(\ref{6.11})$--$(\ref{6.15})$
\begin{equation}
\label{6.21} II
=\bigg\{2[n(\varepsilon)-n(\tilde{\varepsilon})]\langle
S^zS^z\rangle_{\bq} [\oplus]_{\bq}\bigg\}\delta(\omega) .
\end{equation}
Using expression $(\ref{6.16})$ we can unite $(\ref{6.21})$ and
$(\ref{6.20})$
\begin{equation}
\label{6.22} I+II=
\Bigg\{2(n(\varepsilon)-n(\tilde{\varepsilon}))\langle S^z n\rangle_{\bq}
+\frac{1}{2}\left(
\frac{\langle P^+\rangle}{\cosh^2\frac{\beta\varepsilon}{2}}+
\frac{\langle P^-\rangle}{\cosh^2\frac{\beta\tilde{\varepsilon}}{2}}
\right)\Bigg\}\delta(\omega) .
\end{equation}
The diagram series for the fourth term in $(\ref{6.17})$ has the form
$$
\epsfxsize 1.\textwidth\epsfbox{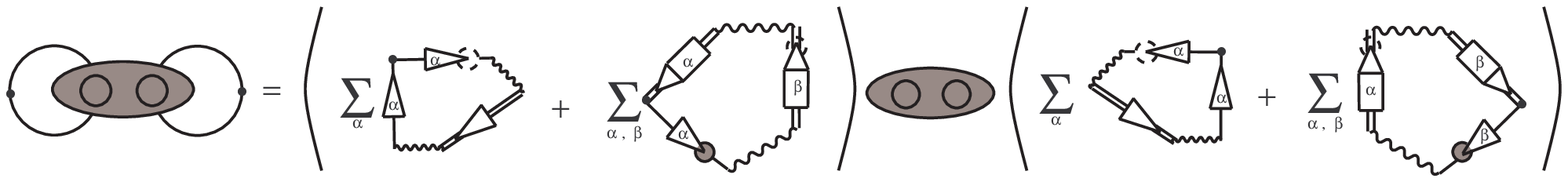}
$$
and can be written as
\begin{equation}
\label{6.23}
IV=
\raisebox{-.6cm}{\epsfysize 1.5cm\epsfbox{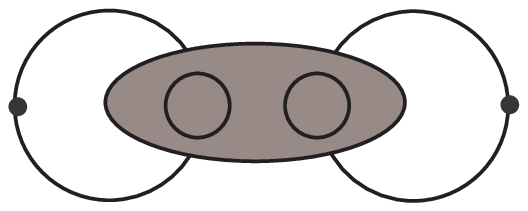}}
=[\oplus]_{\bq} \langle S^z
S^z\rangle_{\bq} [\oplus]_{\bq}\delta(\omega) .
\end{equation}
Using formula $(\ref{6.16})$ once more we unite the $III$-term and the
$IV$-term
\begin{equation}
\label{6.24}
III+IV=\langle n S^z\rangle_{\bq} [\oplus]_{\bq}
\delta(\omega) .
\end{equation}
The last term can be presented in the form
\begin{equation}
{\epsfysize 2.3cm\epsfbox{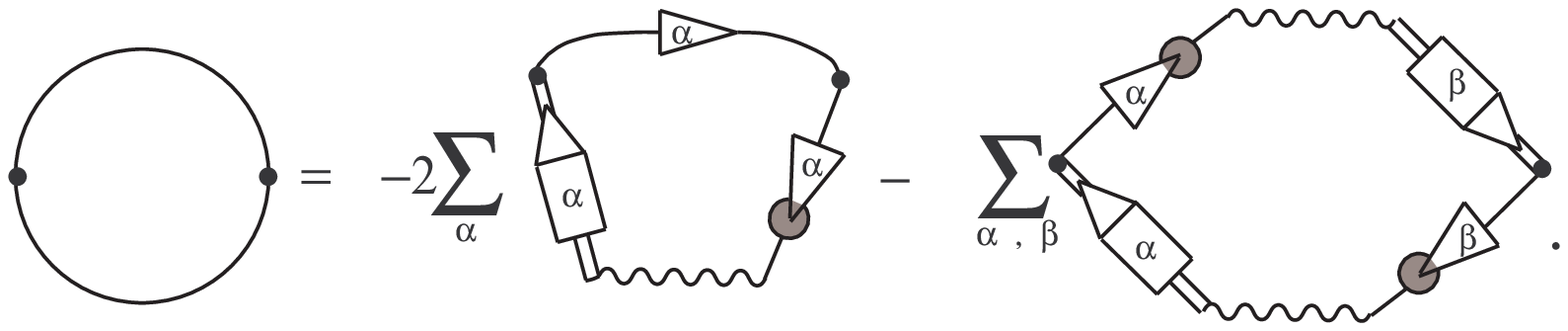}}
\end{equation}
Let us write down the final formula for an electron correlator for the
 uniform (${\bq}=0$) and static ($\omega =0$) case
\begin{eqnarray}
\langle nn \rangle
&\ff=\ff&
2(n(\varepsilon)-n(\tilde{\varepsilon}))\langle S^z n\rangle_{{\bq}=0}+
\frac{1}{2}\left(
\frac{\langle P^+\rangle}{\cosh^2\frac{\beta\varepsilon}{2}}+
\frac{\langle P^-\rangle}{\cosh^2\frac{\beta\tilde{\varepsilon}}{2}}
\right)  \\
\label{6.25}
&&+\frac{\beta}{2N}\sum_{\bk}\frac
{t_{\bk}(\varepsilon-\tilde{\varepsilon})}
{\varepsilon_{\rs I}(t_{\bk})
-\varepsilon_{\rs II}(t_{\bk})}
\left\{ \frac{1}{\cosh^2\frac{\beta\varepsilon_{\rs II}(t_{\bk})}{2}}
-\frac{1}{\cosh^2\frac{\beta\varepsilon_{\rs I}(t_{\bk})}{2}}
\right\}\langle S^z n\rangle_{{\bq}=0}
\nonumber \\
&&+\frac{1}{2N}\sum_{\bk}
\left\{ \frac{1}{\cosh^2\frac{\beta\varepsilon_{\rs II}(t_{\bk})}{2}}
+\frac{1}{\cosh^2\frac{\beta\varepsilon_{\rs I}(t_{\bk})}{2}}
\right\}-\frac{1}{2}
\left\{ \frac{1}{\cosh^2\frac{\beta\varepsilon}{2}}
+\frac{1}{\cosh^2\frac{\beta\tilde{\varepsilon}}{2}}
\right\} .\nonumber
\end{eqnarray}
The same result can be obtained from the derivative
$ {\rm d}\langle n \rangle/({\rm d}\beta\mu).$
Thus, all our quantities: the mean values of the pseudospin and particle
number operators, the thermodynamic potential as well as correlation
functions are derived within the framework of one approximation which
corresponds to the mean field approximation.

\section{Numerical research in the $\boldsymbol{\mu}={}$const regime}

In the investigation of equilibrium conditions we shall separate two
re\-gi\-mes:
$\mu$=const and $n$=const. For the first regime the equilibrium is defined
by the minimum of the thermodynamic potential:
$
\bigg(\frac{\displaystyle\partial\Omega}{\displaystyle\partial\langle S^z
\rangle }\bigg)_{T,\mu ,h}=0
$.

The $\mu$=const regime corresponds to the case when charge
redistribution between the conducting sheets CuO$_2$ and other structural
elements (charge reservoir, e.g., nonstoichiometric in oxygen CuO chains in
YBaCuO-type structures) is allowed.

The dependencies of the order parameter $\langle S^z\rangle$ on field $h$ and
temperature $T$ at the constant value of the chemical potential are
determined by equation (\ref{3.15}). All the integrals in (\ref{3.15}) can be
calculated analytically at zero temperature (below, all the calculations will
be performed for the rectangular density of states, but we would like to
note that the similar behaviour can be obtained in the case of the
semi-elliptic density of states).

We shall present all our results for the case of zero
temperature as well as for the case of non-zero temperature.

\begin{figure}[htbp]
\begin{center}
a)
\raisebox{-4.cm}{\epsfysize 6cm\epsfbox{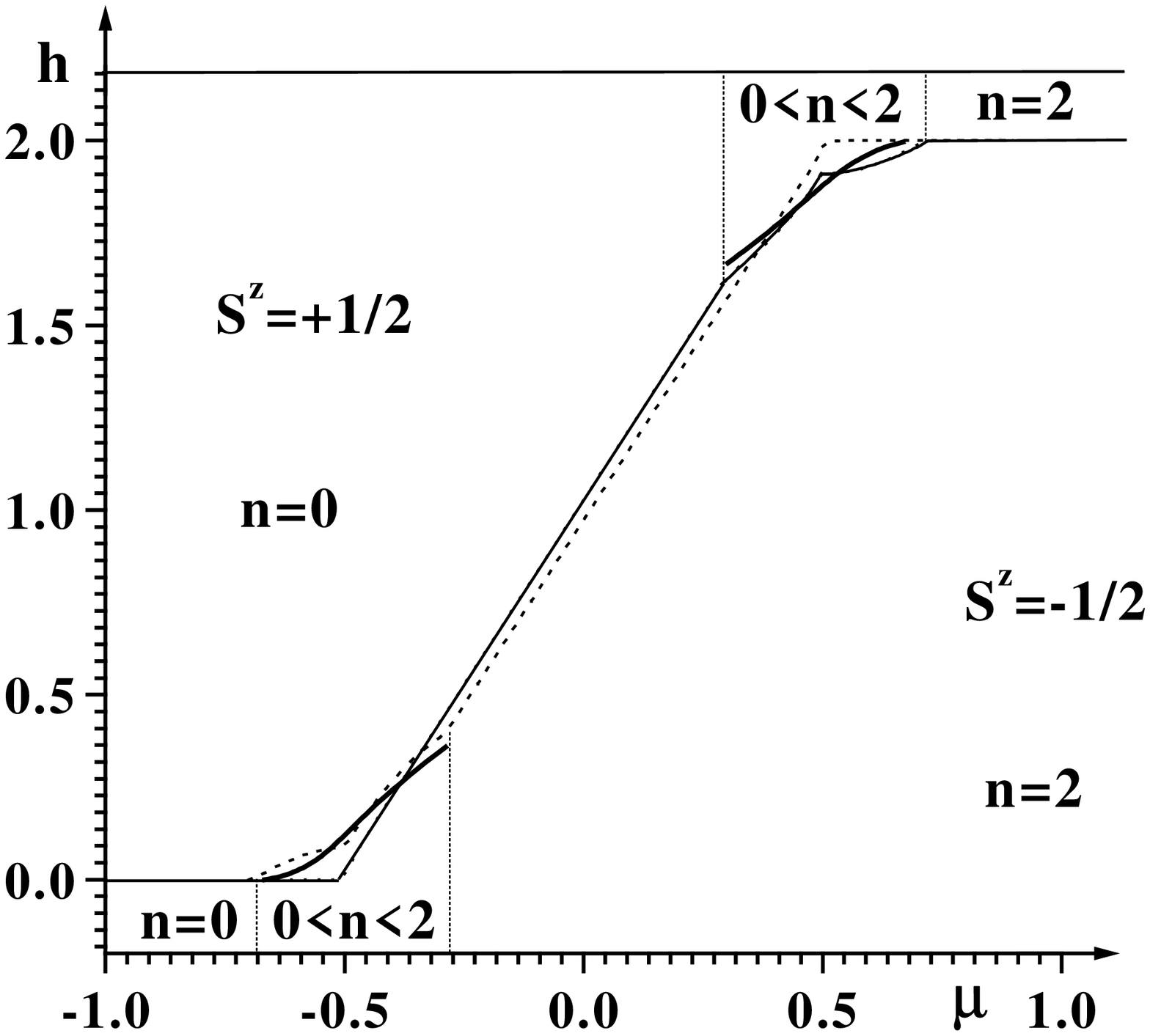}}\quad
b)
\raisebox{-4.cm}{\epsfysize 6cm\epsfbox{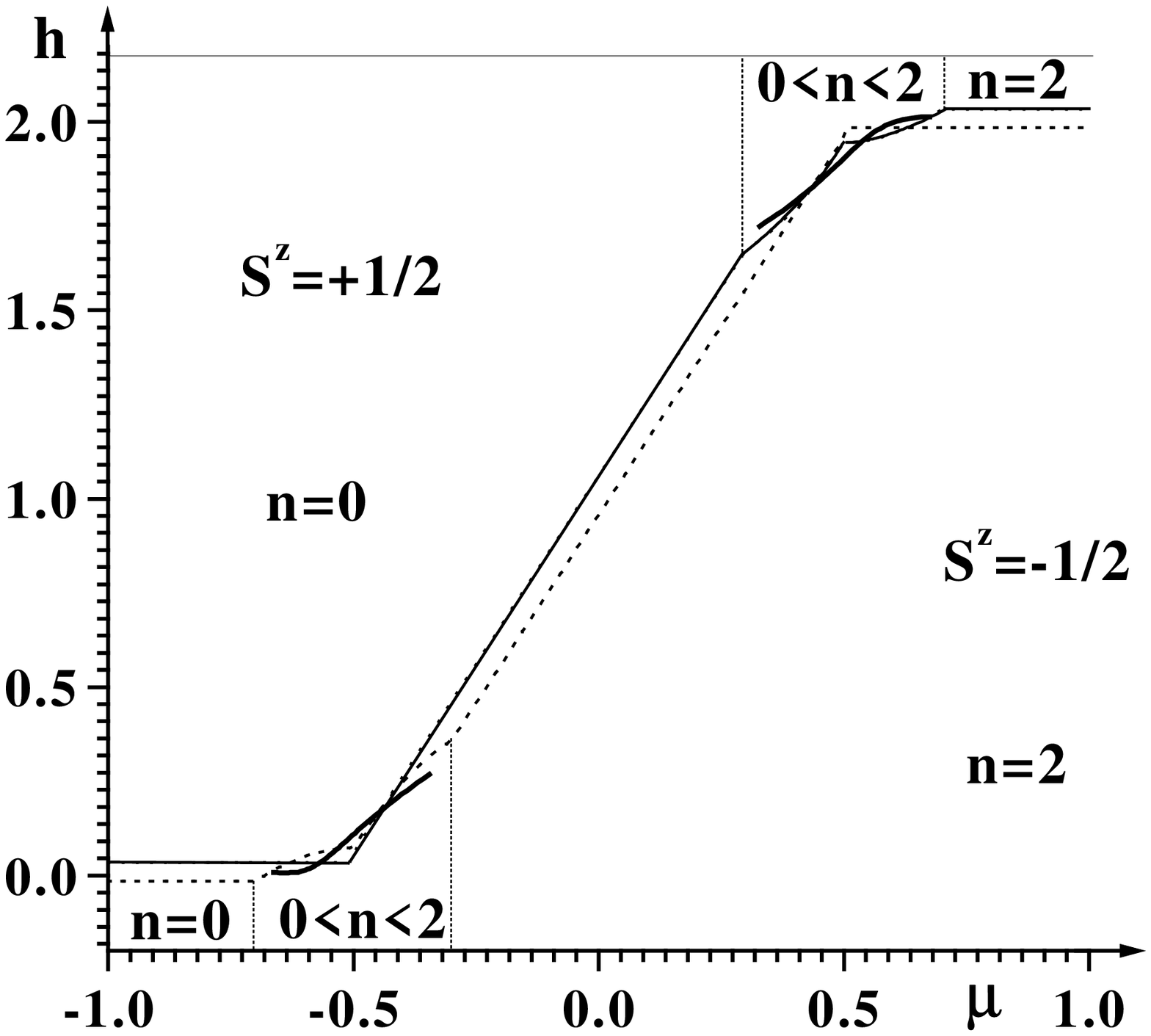}}
\end{center}
\caption{Phase diagram $\mu$ - $h$. Dotted and thin solid lines surround
regions with $S^z=\pm\frac{1}{2}$, respectively.
Thick solid
line indicate the first order phase transition points. a) the case of zero
temperature; b) $T=0.002$ ($W~=~0.2$; $g~=~1$)~.}
\label{2n}
\end{figure}

\begin{figure}[htbp]
\begin{center}
{\epsfysize 6cm\epsfbox{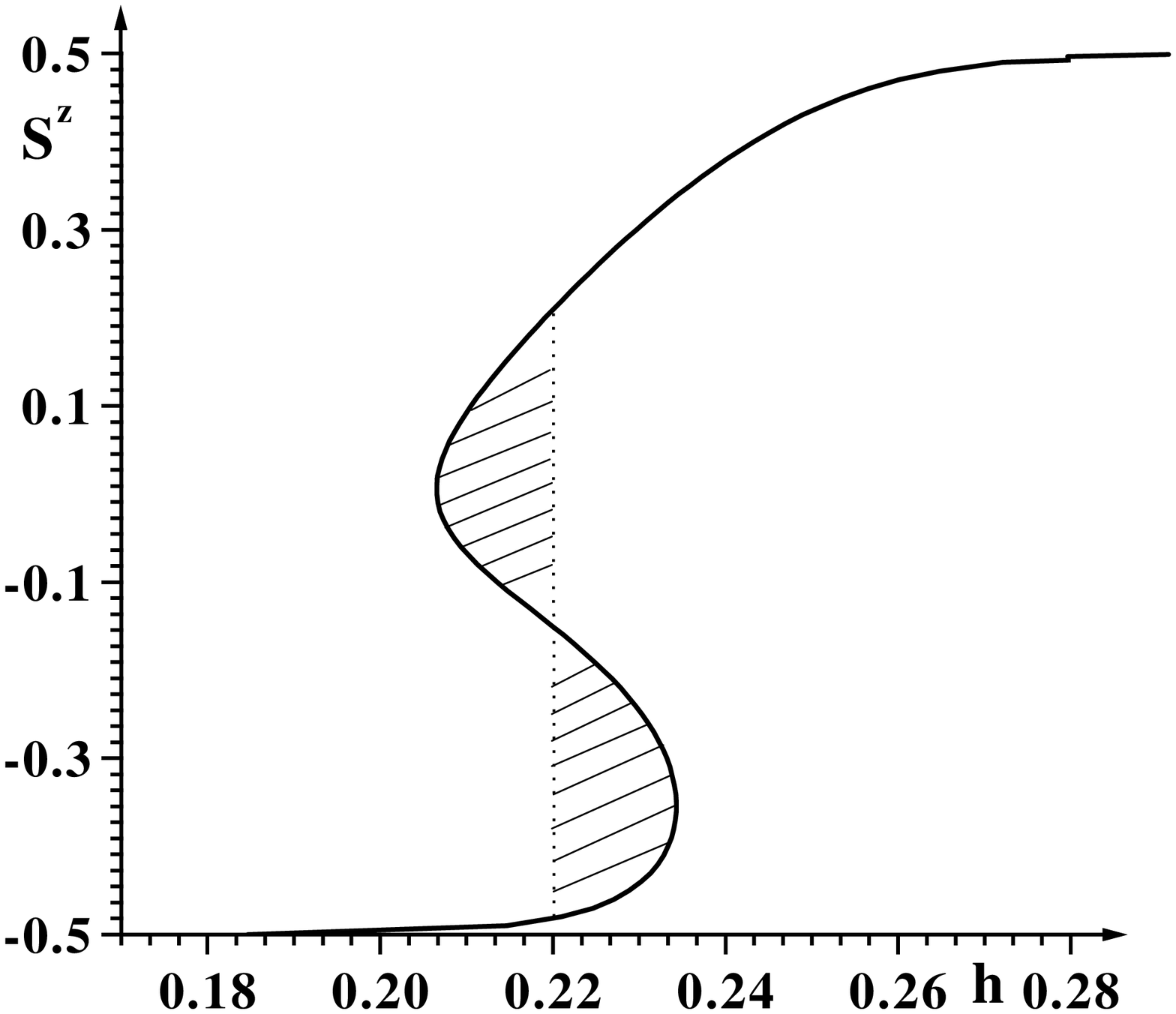}}
\qquad
{\epsfysize 6cm\epsfbox{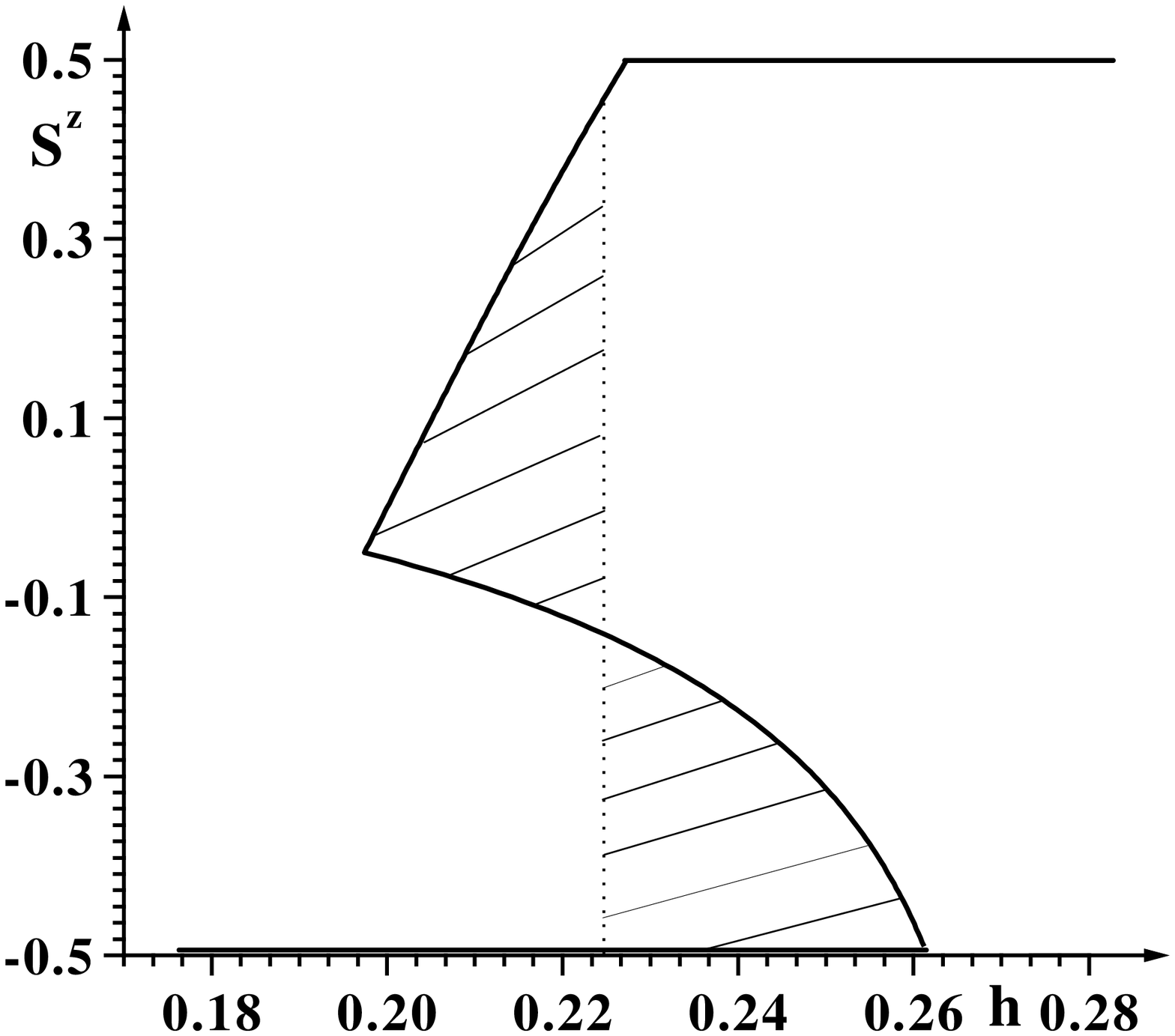}}
\end{center}
\caption{Field dependency of $\langle S^z \rangle $ ($W=0.2$, $\mu
=-0.4$, $g=1.0$) for $\mu$=const regime;  $T=0.01$ and $T=0$.}
\label{3n}
\end{figure}

The phase diagrams $\mu$-$h$ which indicate stability regions for states
with \linebreak
$\langle S^z\rangle=\pm\frac{1}{2}$
are shown in figure~\ref{2n}
for $g\gg W$.

One can see
two regions of the $\mu $ and $h$ values where the states with $\langle
S^z\rangle =\frac{1}{2}$ and $\langle
S^z\rangle =-\frac{1}{2}$ are both stable.
In the vicinity of these regions the
phase transitions of the first order with the change of the longitudinal
field $h$ and/or chemical potential $\mu $ take place and they are shown by
thick lines on phase diagrams in figure~\ref{2n}.

\begin{figure}[htbp]
\begin{center}
{\epsfysize 6cm\epsfbox{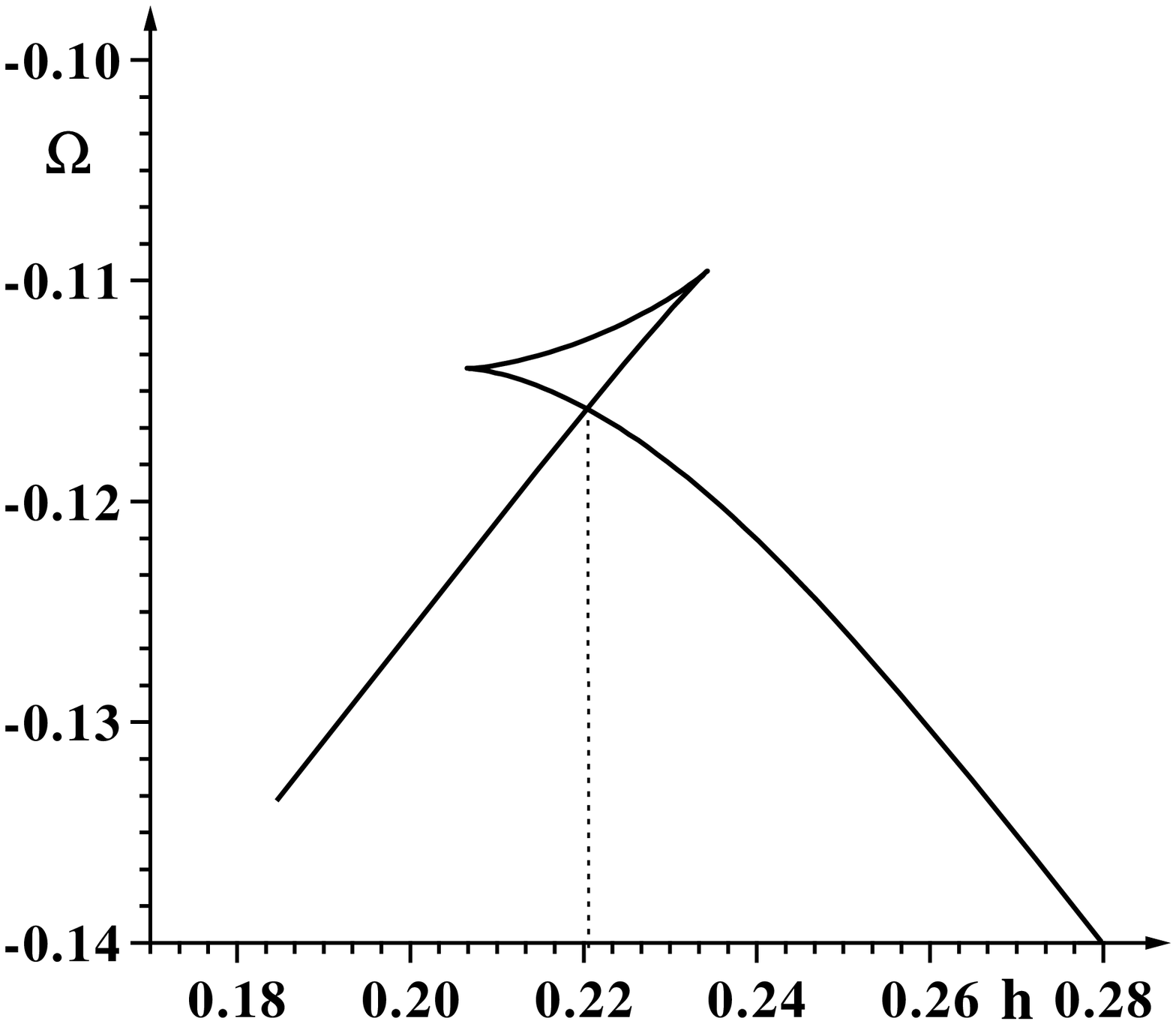}}
\qquad
{\epsfysize 6cm\epsfbox{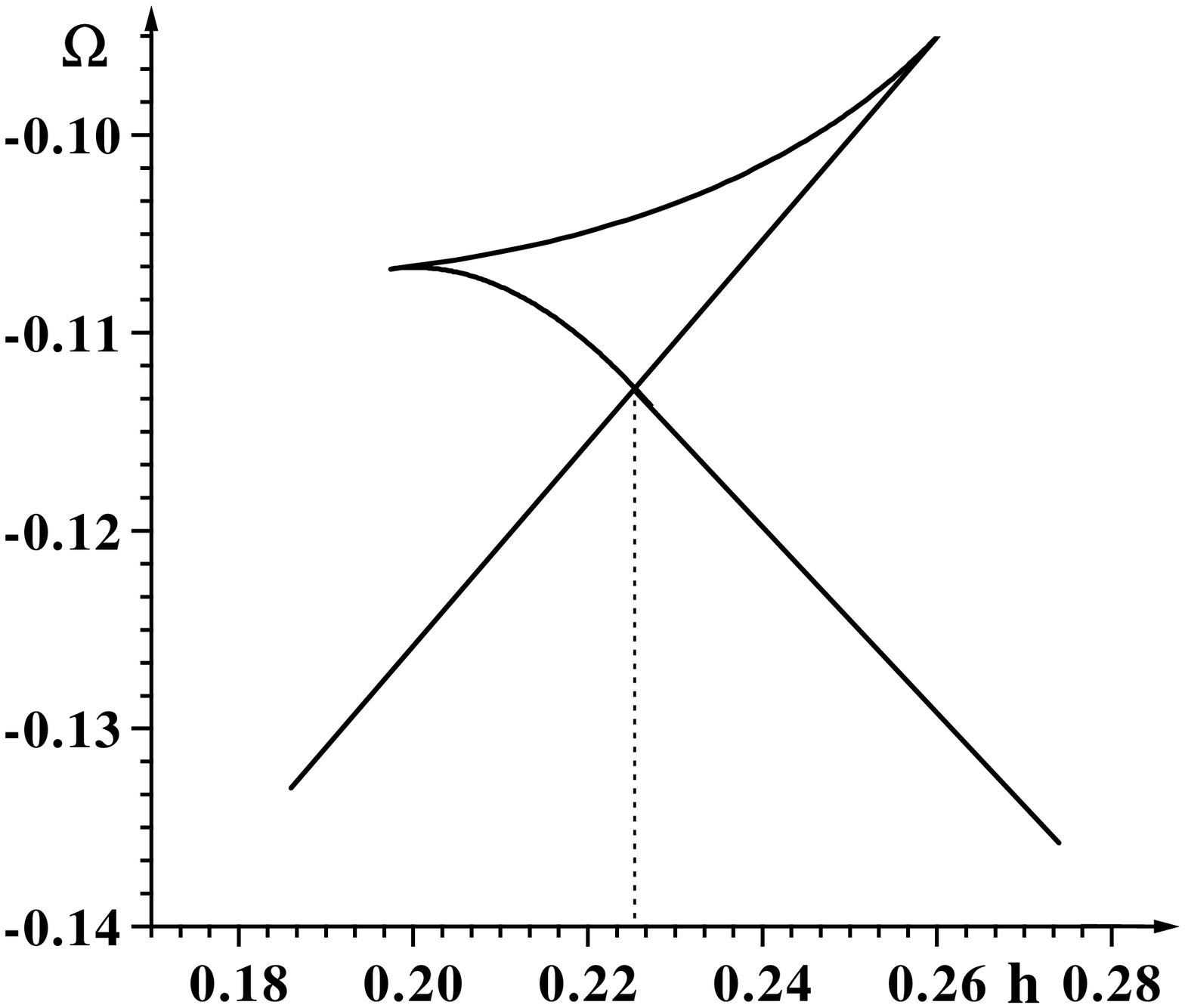}}
\end{center}
\caption{Field dependency of thermodynamic potential ($W=0.2$, $\mu
=-0.4$, $g=1.0$); $T=0.01$ and $T=0$.}
\label{4n}
\end{figure}

The field dependencies of $\langle S^z\rangle$ and $\Omega$
near the phase
transition point are presented in figures~\ref{3n}, \ref{4n}. Their
behaviour in the cases of $T=0$ and $T\neq 0$ with the change of the chemical
potential is similar: S-like for the mean value of the pseudospin and
a fish tail form for the thermodynamic potential.

\begin{figure}[htbp]
\begin{center}
{\epsfysize 6cm\epsfbox{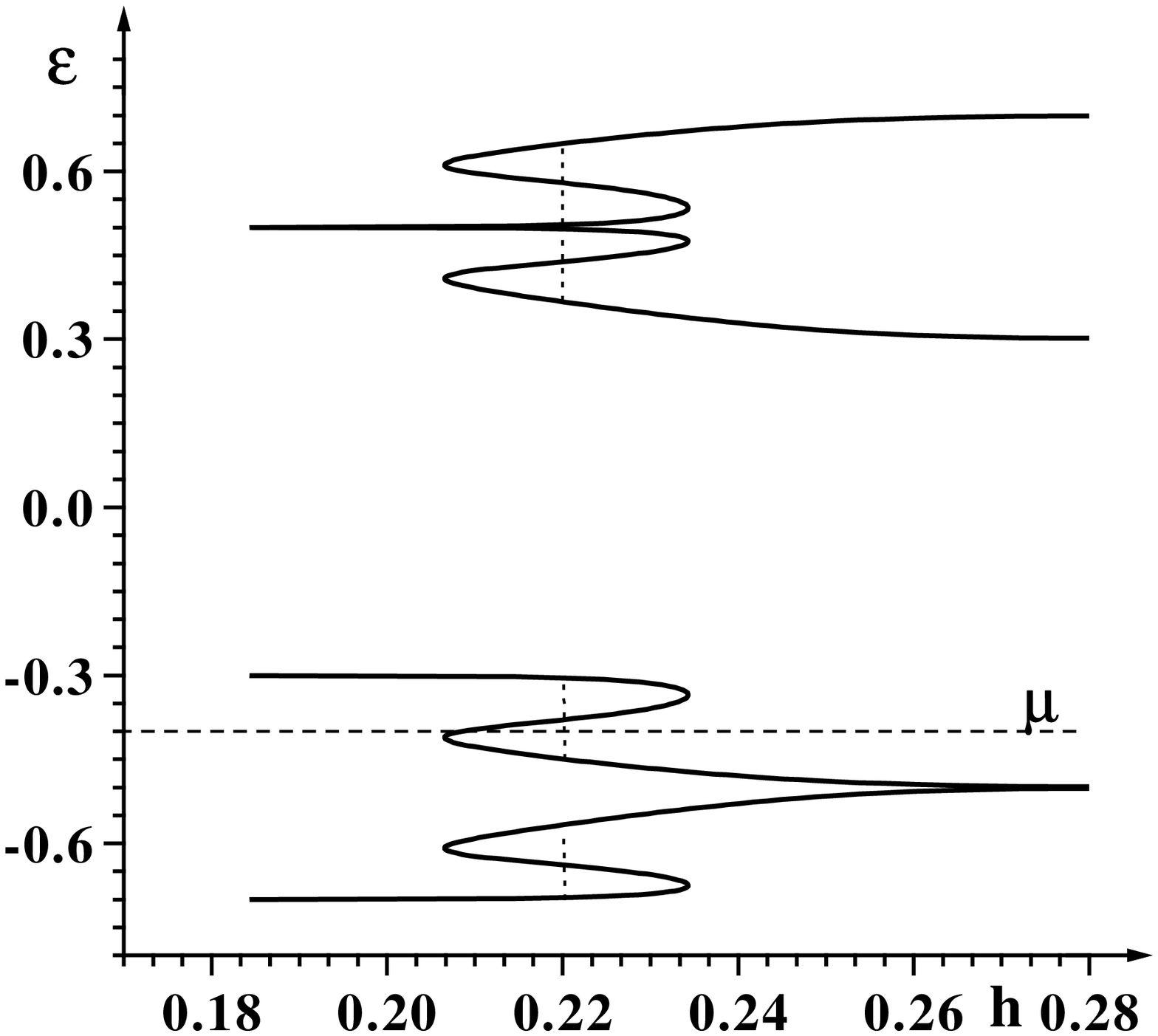}}
\qquad
{\epsfysize 6cm\epsfbox{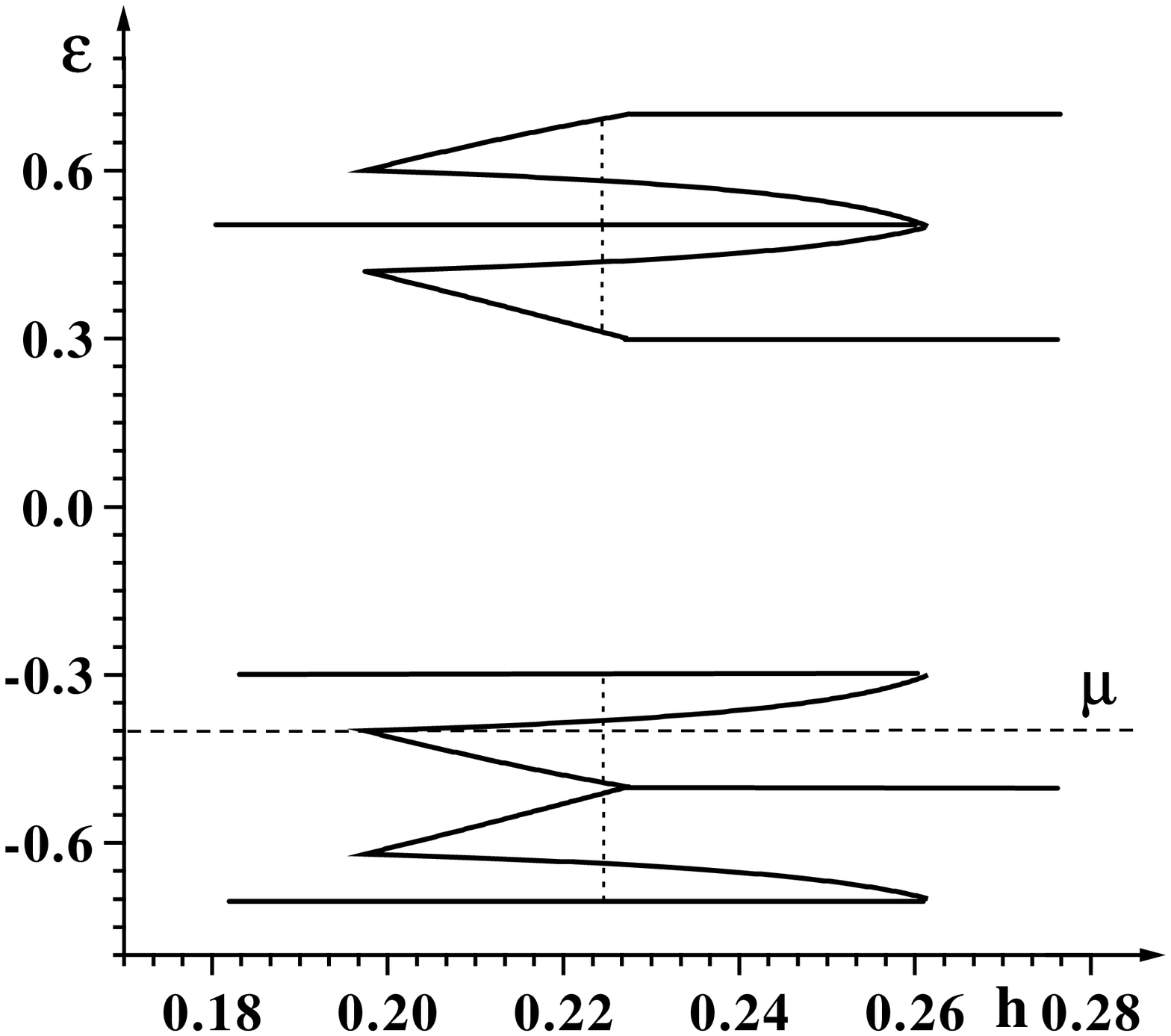}}
\end{center}
\caption{Field dependency of electron bands boundaries ($W=0.2$, $\mu
=-0.4$, $g=1.0$); $T=0.01$ and $T=0$.}
\label{5n}
\end{figure}

\begin{figure}[htbp]
\begin{center}
{\epsfysize 6cm\epsfbox{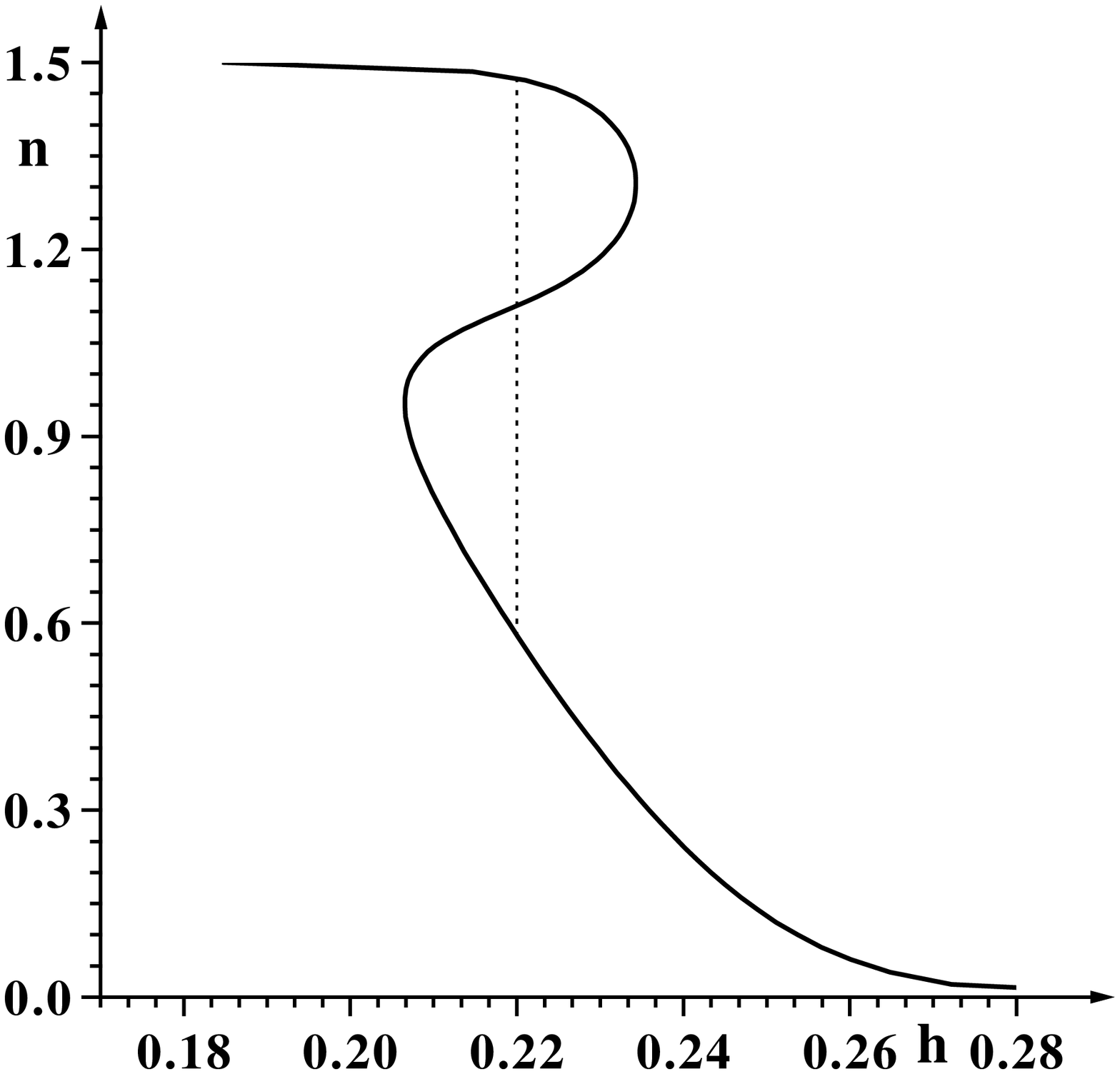}}
\qquad
{\epsfysize 6cm\epsfbox{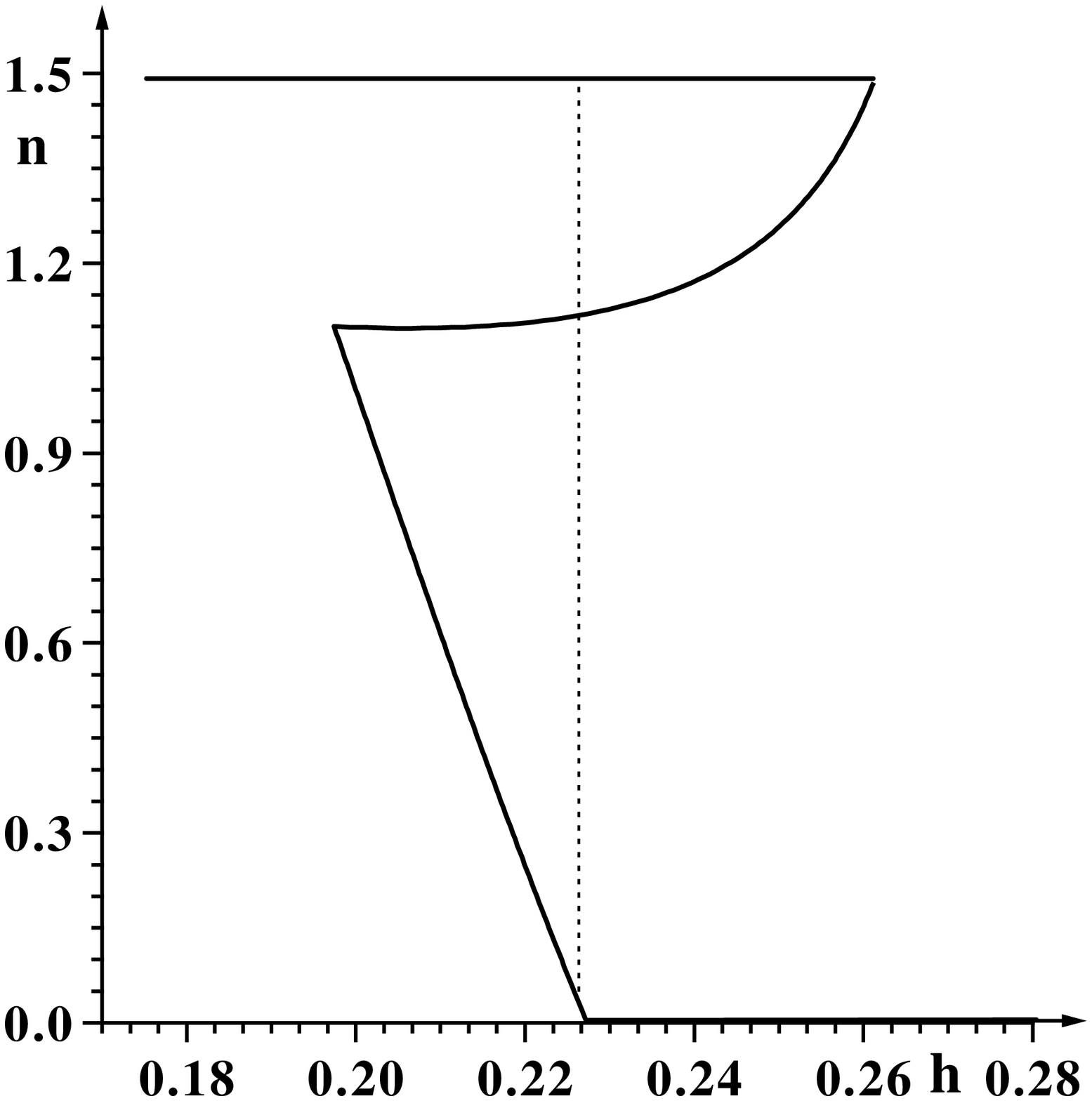}}
\end{center}
\caption{Field dependency of electron concentration ($W=0.2$,
$\mu~=~-0.4$, $g=1.0$); $T=0.01$ and $T=0$.}
\label{6n}
\end{figure}

In the $\mu$=const regime, the chemical potential can appear
in the electron bands or out of them with the change of field
$h$ (dashed line in figure~\ref{5n}),
and in the vicinity of the phase transition
point this results in a rapid change of electron
concentration (dotted lines in figure~\ref{6n})
due to a charge transfer from/to
the reservoir (CuO planes).
The widths of the electron subbands depend on the mean value of the
pseudospin which results in the presented
above behaviours.

The phase transition point is presented by a crossing
point on the dependence
$\Omega (h)$ (figure~\ref{4n}).
At the same time this value is determined according to the Maxwell rule
from the plot of function $S^z(h)$.

\begin{figure}[htbp]
\begin{center}
a)\raisebox{-4.cm}{\epsfysize 6cm\epsfbox{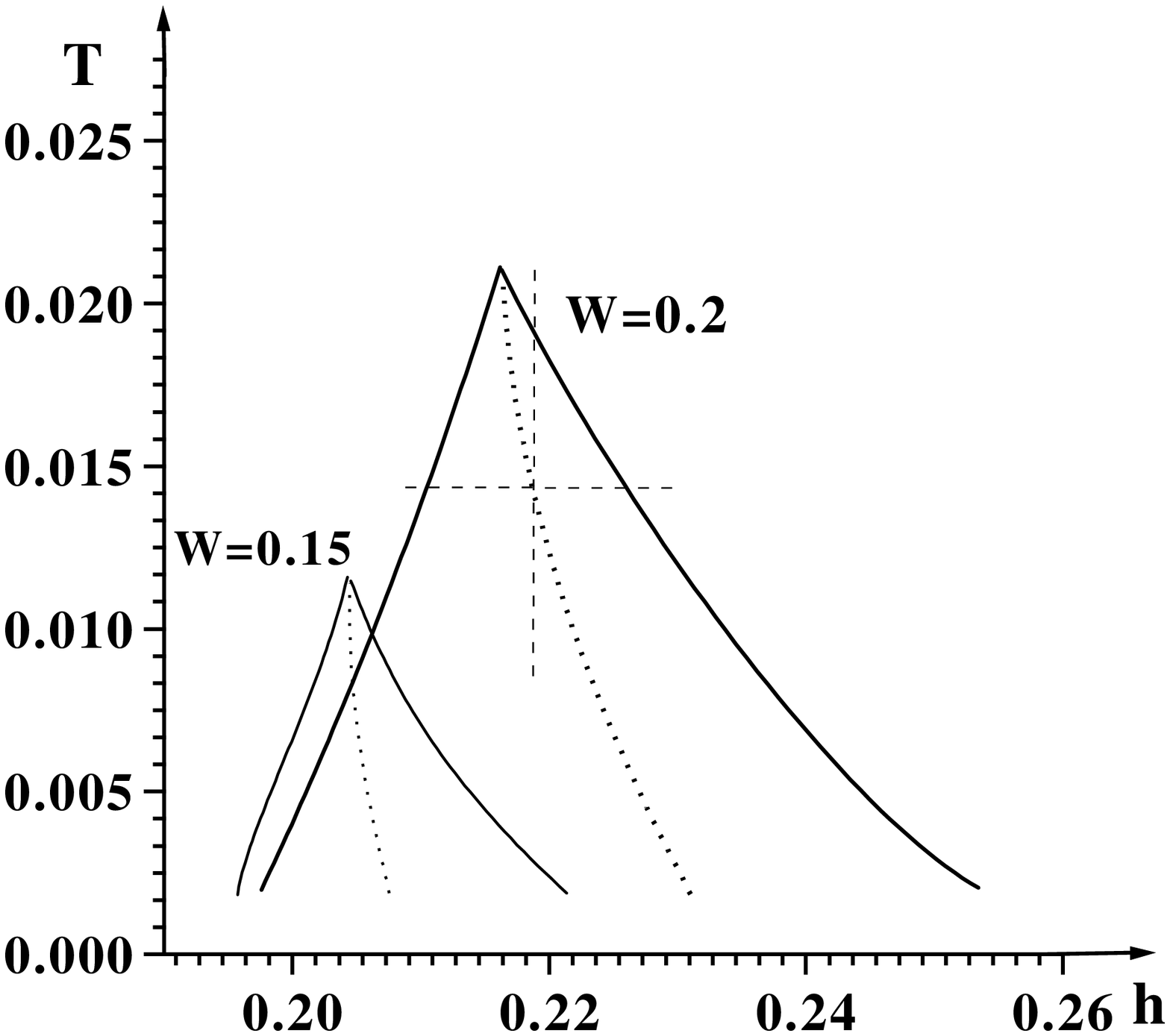}}
b)\raisebox{-4.cm}{\epsfysize 6cm\epsfbox{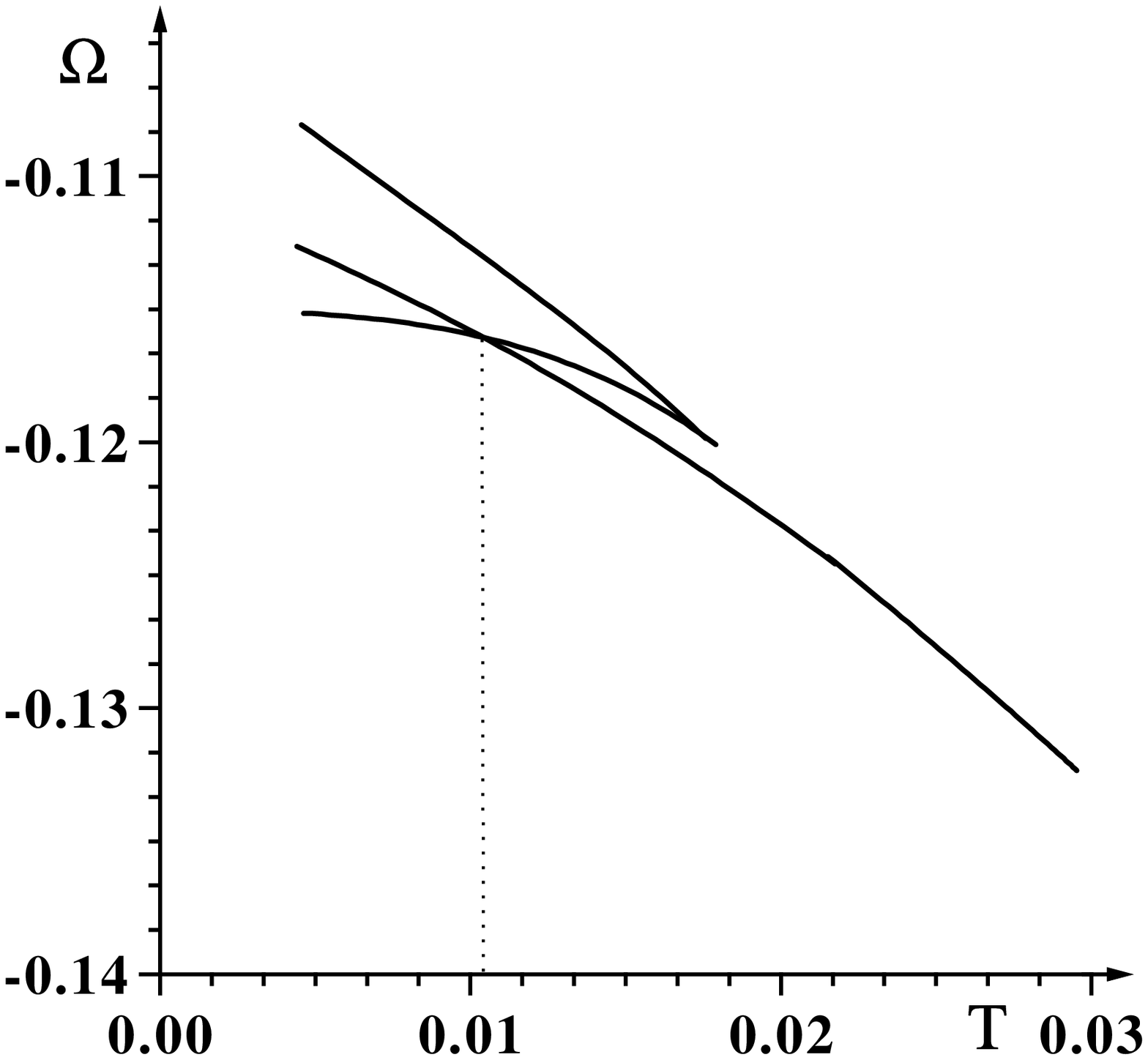}}
\end{center}
\caption{a) phase diagram $T_{\rm c}$-$h$ ($g=1$, $\mu =-0.4$);\quad
b) temperature dependence of the thermodynamic potential
($W~=~0.2$, $h=0.22$, $\mu=-0.4$, $g=1$).}
\label{7n}
\end{figure}

With the temperature increase the region of phase coexistence narrows,
and the corresponding phase diagram $T_c$-$h$ is shown in
figure~\ref{7n}a.
The tilt of the coexistence curve testifies to the
possibility of the first order phase transition at a change of
temperature with a jump of the pseudospin mean value.
(The phase diagram $T_c$-$\mu$ has a similar form).
The existence of the shifted and tilted
coexistence curve as the result of the local
pseudospin-electron interaction was obtained for the first time
in \cite{12} for a pseudospin-electron
model with a direct interaction between pseudospins.

\begin{figure}[htbp]
\begin{center}
{\epsfysize 6cm\epsfbox{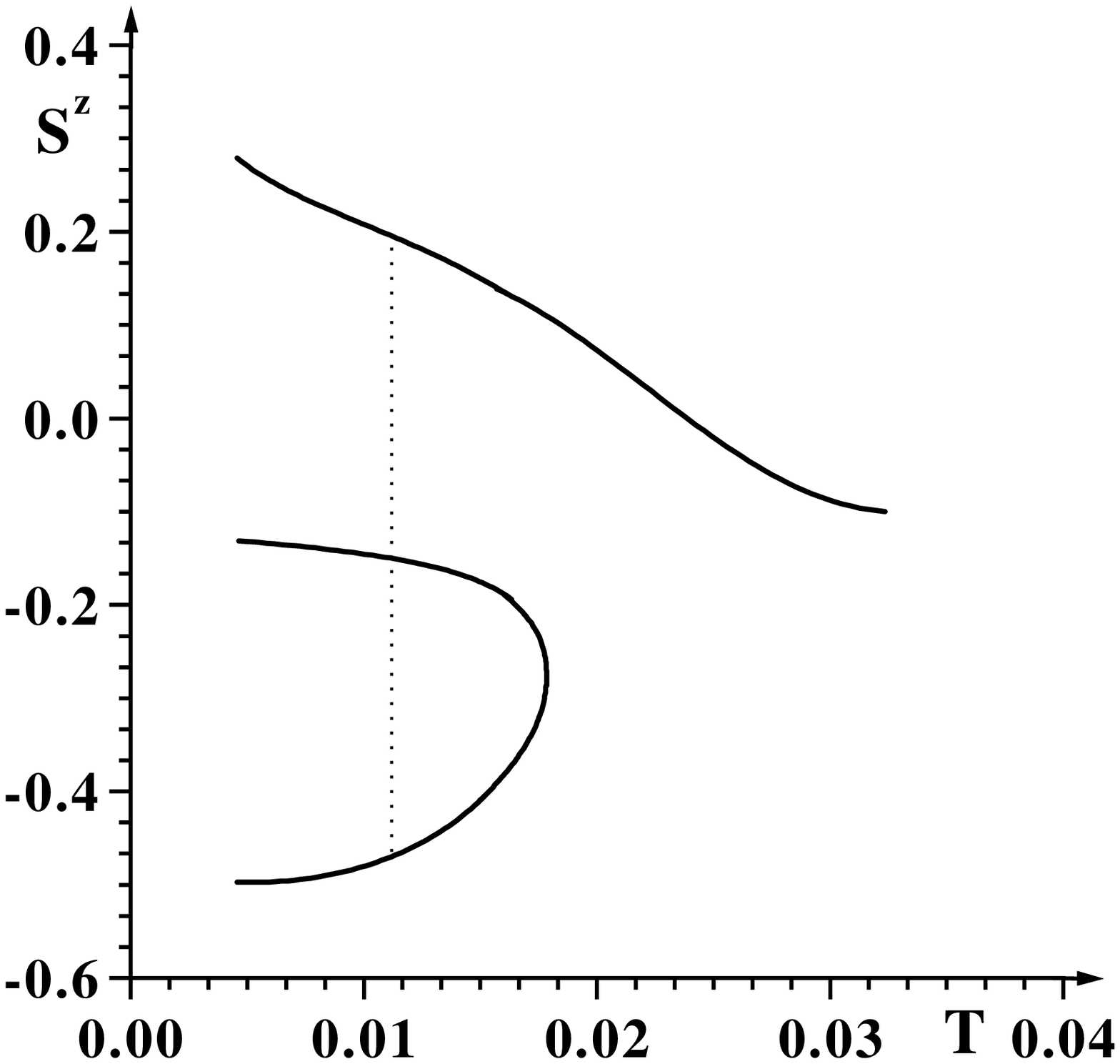}}
\qquad
{\epsfysize 6cm\epsfbox{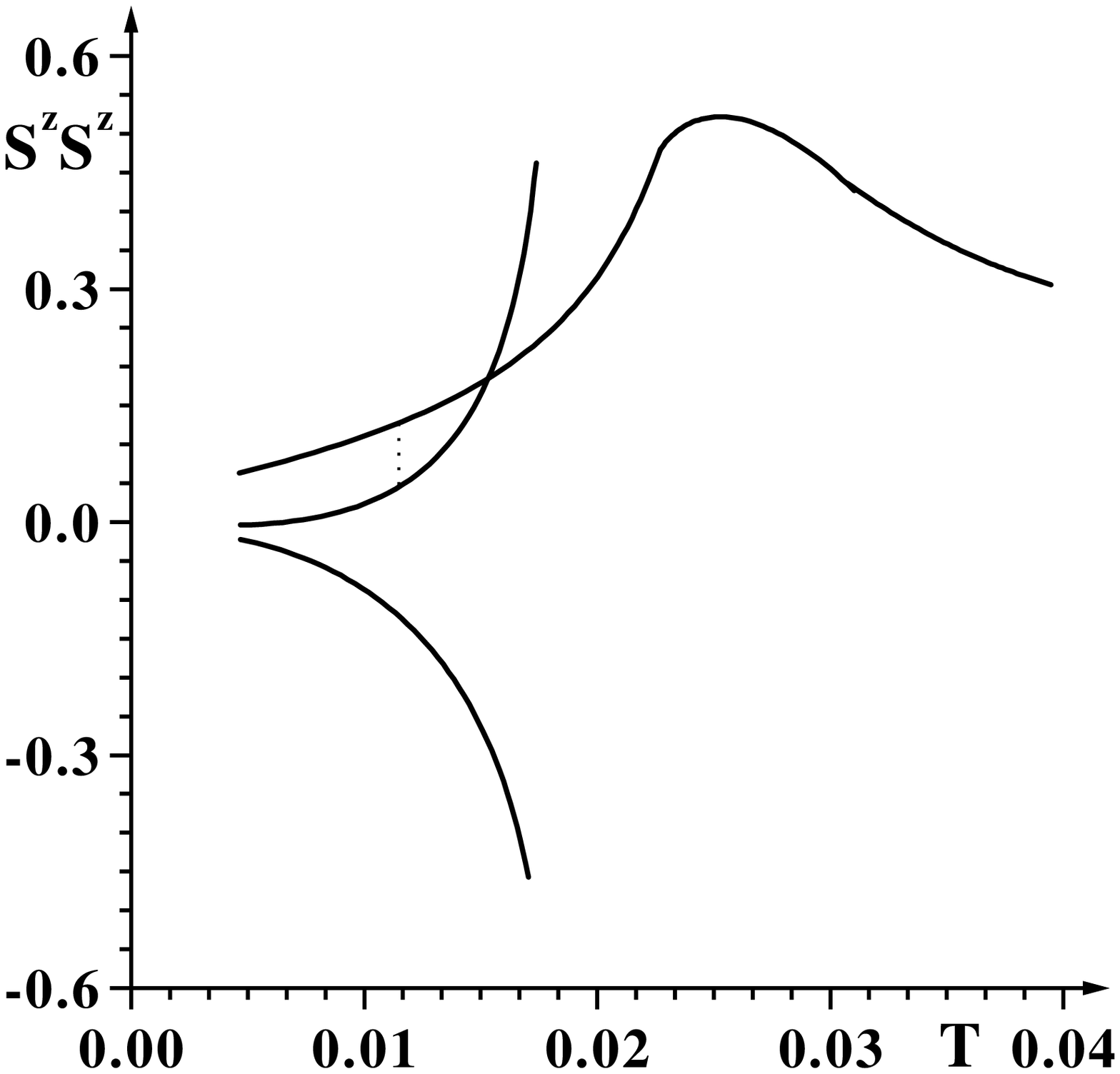}}
\end{center}
\caption{Dependence of the mean value of the pseudospin and
the pseudospin-pseudospin correlation function
on temperature ($W=0.2$, $h=0.22$, $\mu~=~-0.4$, $g=1$).}
\label{15n}
\end{figure}

The analysis of the thermodynamic potential behaviour
with the  temperature increase
(figure~\ref{7n}b) shows that the lowest value of $\Omega (T)$ corresponds
to the jump of the  mean value of the pseudospin
(dotted lines in figure~\ref{15n})
from the branch
which corresponds to the low temperature phase to that of the hight
temperature phase.
The analysis of the
$\langle S^zS^z\rangle$ behaviour with the temperature decrease shows that
the high temperature phase is stable up to zero temperature.
This means that the vertical line on
the $T_{\rm c}$-$h$ phase diagram only once crosses the boundary of the phase
stability.

In figures the case when the chemical potential is placed in the lower
subband is presented. There is no specific behaviour
when the chemical potential is placed out of the bands. And if the chemical
potential is placed in the upper subband our results transform according to
the internal symmetry of the Hamiltonian:
\begin{equation}
\label{7.1}
\mu\to -\mu,\hspace{2em} h\to 2g-h, \hspace{2em} n\to 2-n,\hspace{2em}
S^z\to -S^z.
\end{equation}

\section{Numerical research in the $\boldsymbol{n}={}$const regime}

In the regime of a fixed value of electron concentration the first
order phase transition with a jump of the pseudospin mean value accompanied
by a change of electron concentration
transforms into a phase separation.

\begin{figure}[htbp]
\begin{center}
{\epsfysize 6cm\epsfbox{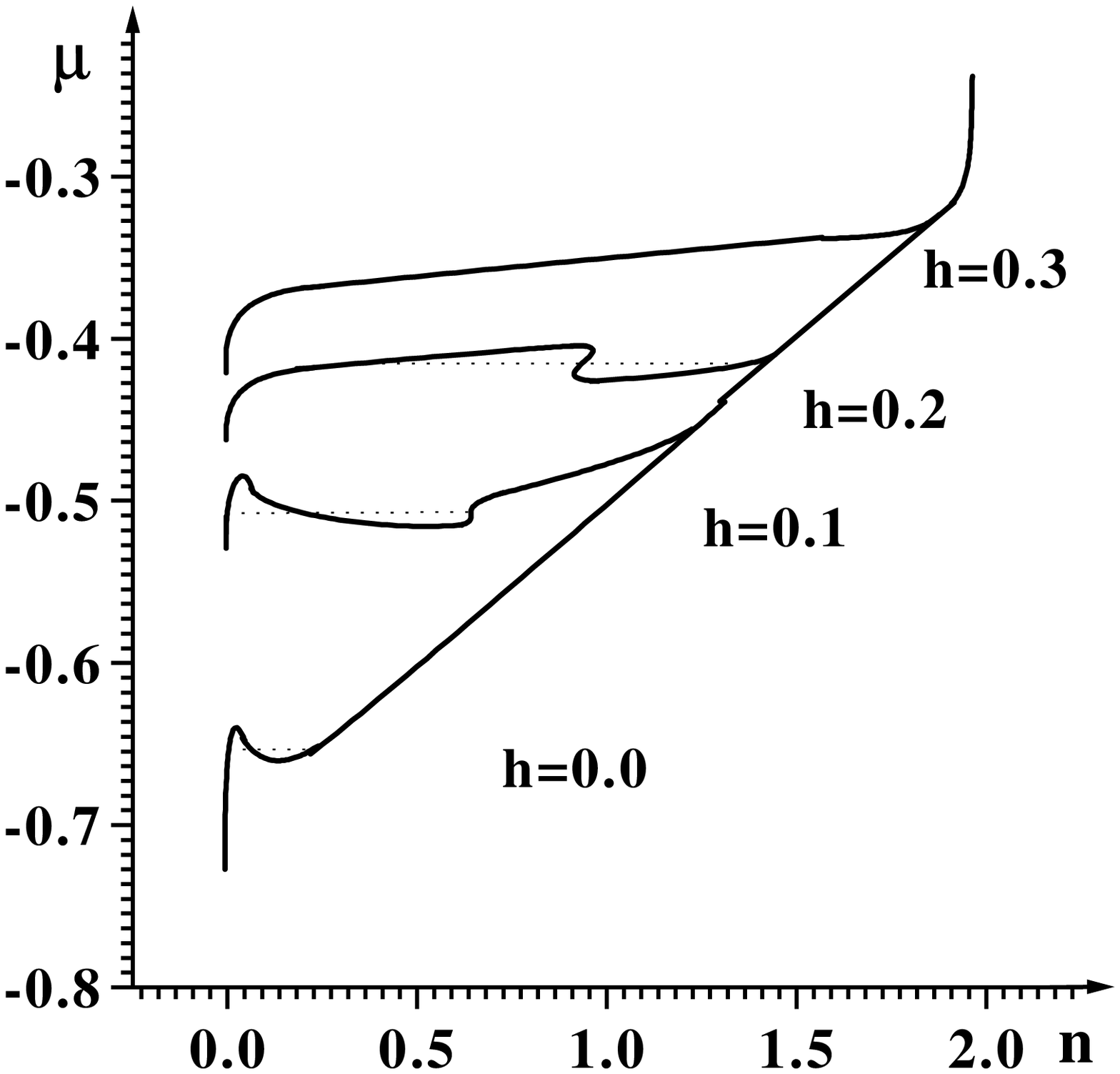}}
\qquad
{\epsfysize 6cm\epsfbox{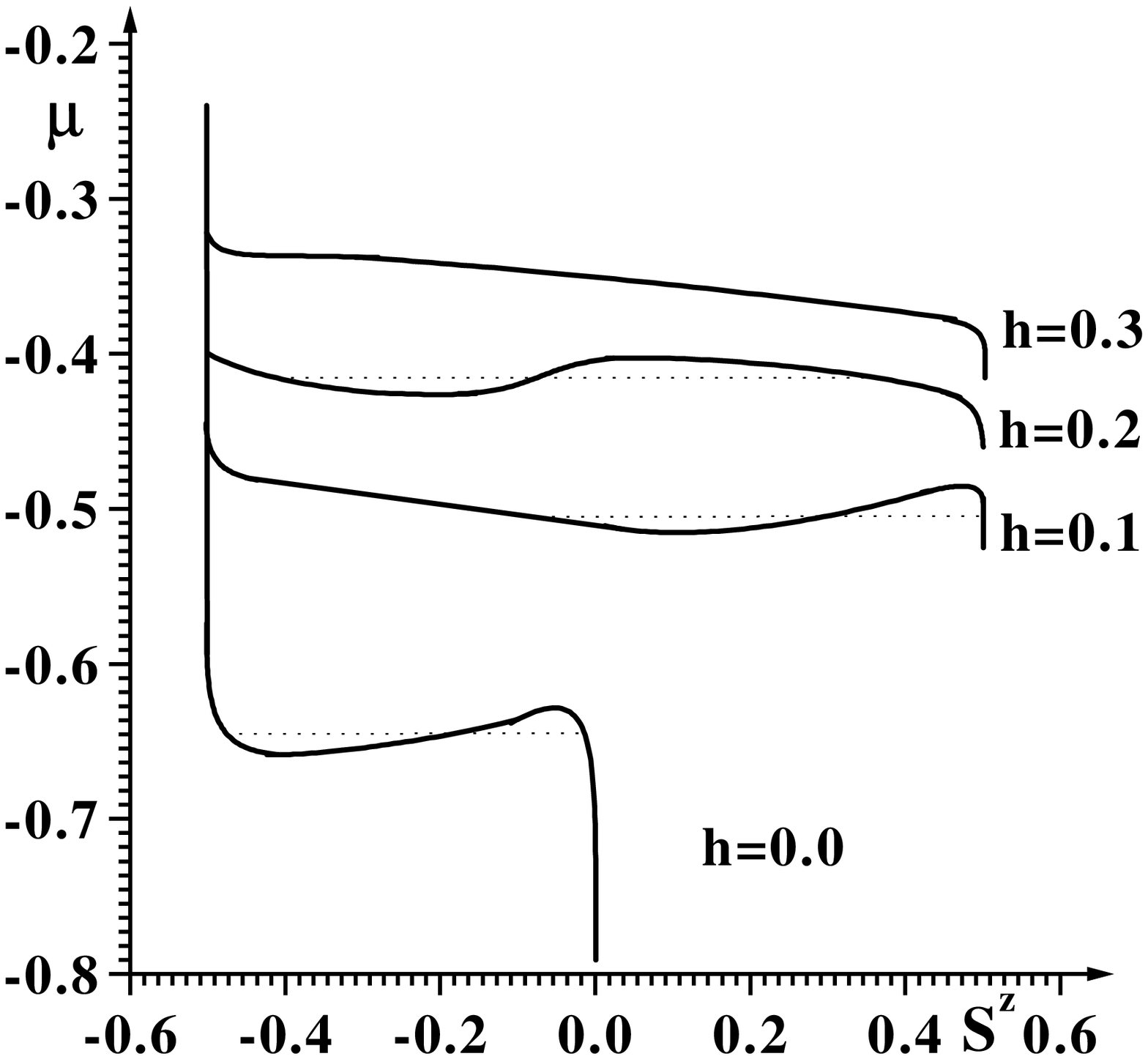}}
\end{center}
\caption{Dependence of the chemical potential $\mu $ on the electron
concentration $n$ and the pseudospin mean value $\langle S^z\rangle $ for
different $h$ values ($g=1$, $W=0.2$, $T=0.01$).}
\label{8n}
\end{figure}

\begin{figure}[htbp]
\begin{center}
{\epsfysize 6cm\epsfbox{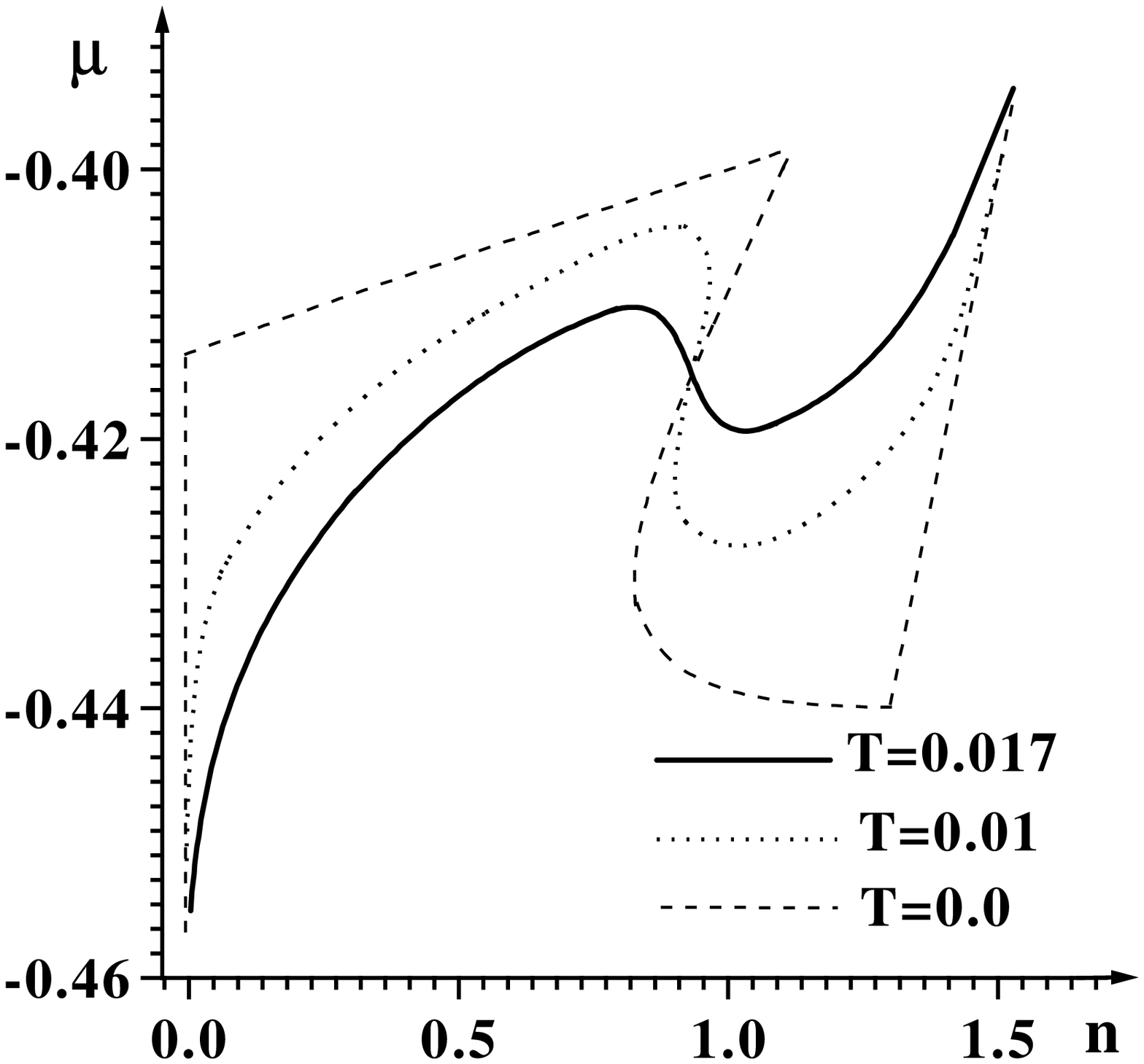}}
\qquad
{\epsfysize 6cm\epsfbox{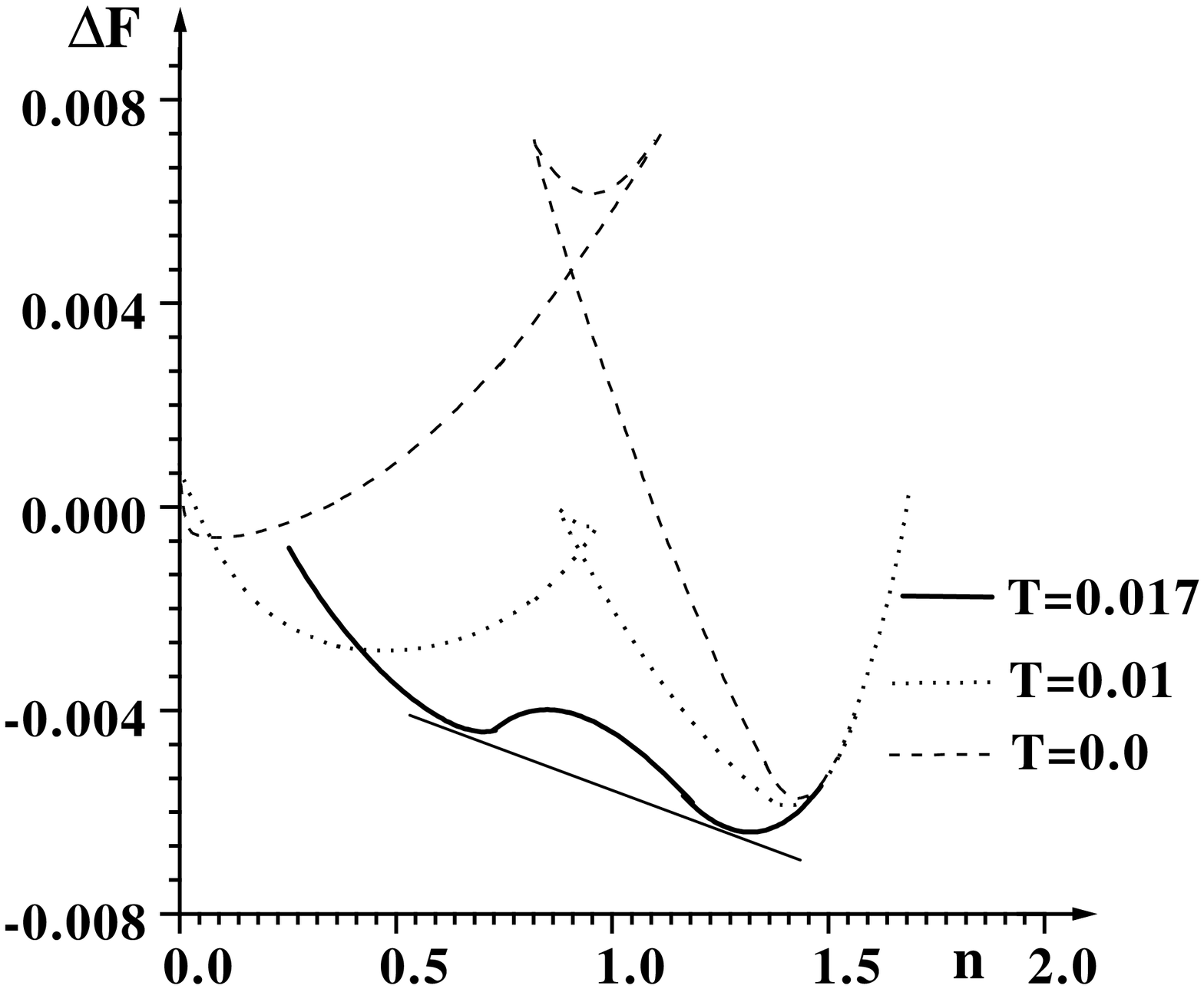}}
\end{center}
\caption{Dependence of the chemical potential $\mu $ on the electron
concentration $n$ and deviation of the free energy from linear dependence for
different $T$ values ($g=1$, $W=0.2$, $h=0.2$).}
\label{13n}
\end{figure}

The dependence of the mean value of the particle number
(or electron concentration) on the chemical potential is one of
the factors determining thermodynamically stable states of the
system. One can see the regions with ${\rm d}\mu /{\rm d}n\leqslant 0$
where states with a homogenous distribution of
particles are unstable, which corresponds to the phase
separation into the regions with different electron concentrations and
pseudospin mean values (figures~\ref{8n} and \ref{13n}).

In the $n$=const regime the equilibrium condition is determined by the
minimum of free energy $F=\Omega +\mu N$. In the phase
separated region the free energy as a function of $n$
deflects up (figure~\ref{13n})
and concentrations of the separated phases are determined by
the tangent line touch points (these points are also the points of binodal
lines which are determined according to the Maxwell rule from the function
$\mu (n)$, see figure~\ref{8n}).

\begin{figure}[htbp]
\begin{center}
{\epsfysize 8cm\epsfbox{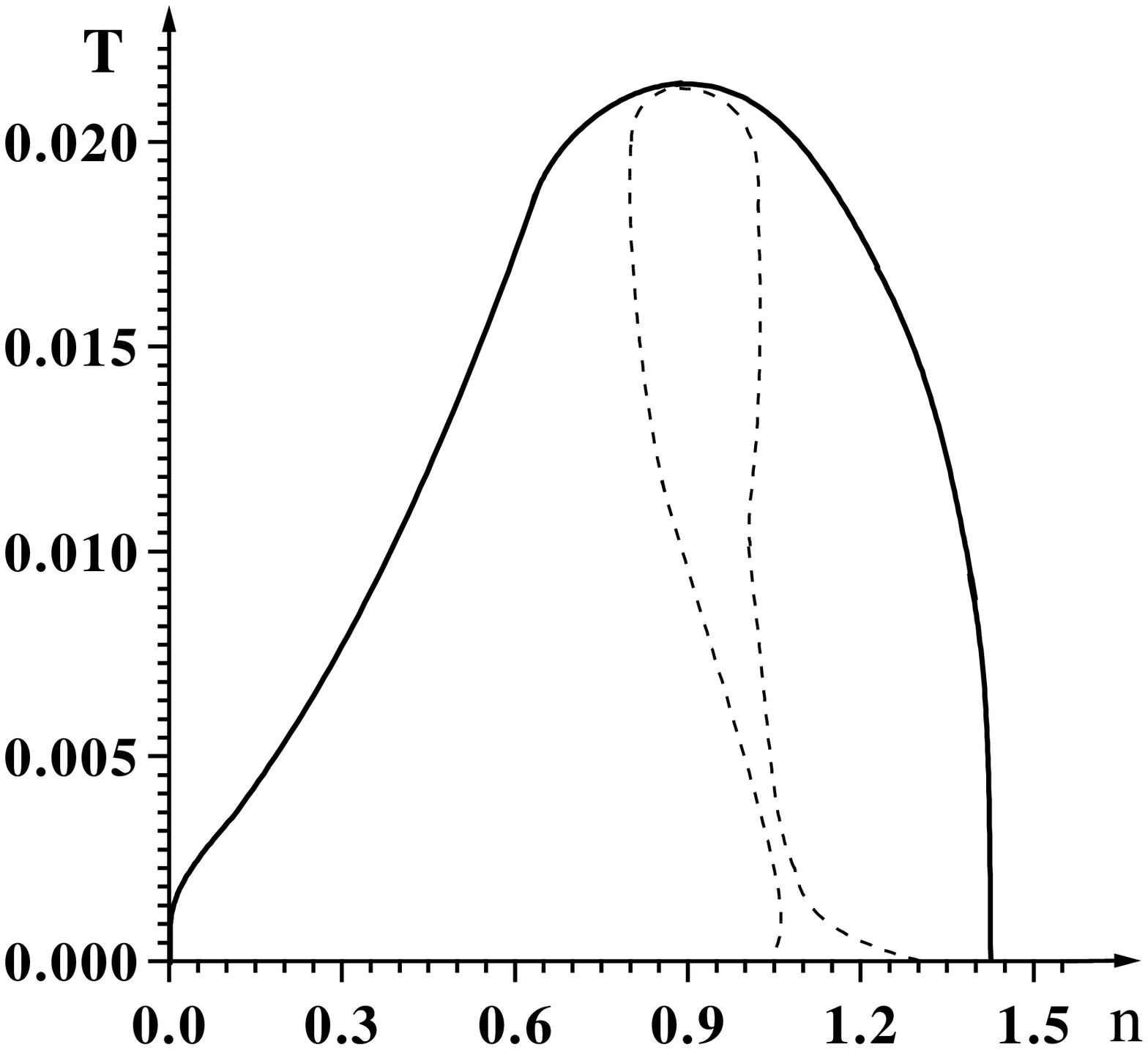}}
\end{center}
\caption{Phase diagram $T$-$n$ ($g=1$, $W=0.2$ $h=0.2$) for a phase
separated state. Solid line indicates binodal points, dashed line
indicates spinodal points.}
\label{14n}
\end{figure}

The resulting phase diagram $T$-$n$ is shown in figure~\ref{14n}.

\section{Conclusions}

Investigation of a pseudospin-electron model in the case of electron
correlation absence was performed in the mean field approximation using the
Hubbard-I
approximation for the calculation of a single-particle Green function.
We presented the analytical consideration of our model and all the
quantities were obtained within the framework of one self-consistent
approximation.

As the result of numerical investigations we have obtained:

1) there is always a gap in the electron spectrum (with a
change of the mean value of the pseudospin a reconstruction of the electron
spectrum takes place);

2) the possibility of the first
order phase transition with a change of the
longitudinal field $h$ (as a consequence of this the $S$-like behaviour of
the mean value of the pseudospin with a jump in the phase transition point
(which corresponds to the inflected point on the  dependence $\Omega (h)$)
is obtained and at this point the concentration rapidly redistributes
between the conducting sheets CuO$_2$ and the charge reservoir (CuO planes)
in YBaCuO-type structures);

3) the phase coexistence curve is tilted from the vertical line, therefore,
there exists a possibility of the first order phase transition with the
temperature change;

4) the high temperature phase
is stable in the whole region of temperatures (figure~\ref{15n});

5) in the regime $n$=const we have the regions with
${\rm d}\mu /{\rm d}n<0$, which
corresponds to phase separation with the appearance of regions with
different electron concentrations and different orientations of pseudospins.

Analytical expressions for the mean values, thermodynamic functions
and susceptibilities of such a simplified pseudospin-electron model
($U=0$, $\Omega =0$) were obtained for the uniform phase, when
$\langle S_i^z\rangle =\langle S^z\rangle$, and only the
possibilities of phase transitions with uniform changes
(${\bq }=0$) were analysed numerically. On the other hand,
it is known that for certain parameter values the
charge ordered phase can exist in a strong
coupling limit of the pseudospin-electron model
($U\to\infty$)~\cite{13}
which calls for the consideration of the  possible superstructure
orderings in the opposite limit of $U=0$ and will be the subject of
our further investigations.

 $$ $$
 Tabunshchyk K.V. e-mail: tkir@icmp.lviv.ua

\begin{thebibliography}{18}

\bibitem{1}  Cohen~R.E., Pickett~W.E., Krakauer~H. Theoretical
determination of strong electron-phonon coupling in YBa$_2$Cu$_3$O$_7$. //
Phys.~Rev.~Lett., 1990, vol.~64, No.~21, \mbox{p.~2575--2578.}
\bibitem{2} Maruyama~H., Ishii~T., Bamba~N., Maeda~H., Koizumi~A.,
Yoshikawa~Y., Yamazaki~H. Temperature dependence of the EXAFS spectrum in
YBa$_2$Cu$_3$O$_{7-\delta}$ compounds. // Physica~C, 1989,
vol.~160, No.~5/6, \mbox{p.~524.}
\bibitem{3} Conradson~S.D., Raistrick~I.D. The axial oxygen atom and
superconductivity in YBa$_2$Cu$_3$O$_7$. // Science, 1989,
vol.~243, No.~4896, \mbox{p.~1340.}
\bibitem{4} M\"uller~K.A. // Phase transition, 1988,
(Special issue).
\bibitem{5} Bishop~A.R., Martin~R.L., M\"uller~K.A., Tesanovic~Z.
Superconductivity in oxides: Toward a unified picture. // Z.~Phys.~B -
Condensed Matter, 1989, vol.~76, No.~1, \mbox{p.~17--24.}
\bibitem{6} Hirsch~J.E., Tang~S. Effective interactions in an oxygen-hole
metal. // Phys.~Rev.~B, 1989, vol.~40, No.~4, \mbox{p.~2179--2186.}
\bibitem{7} Frick~M., von der Linden~W., Morgenstern~I., Raedt~H. Local
anharmonic vibrations, strong correlations and superconductivity: A quantum
simulation study. // Z.~Phys.~B - Condensed Matter, 1990,
vol.~81, No.~2, \mbox{p.~327--335.}
\bibitem{8} Stasyuk~I.V., Shvaika~A.M.
Dielectric properties and electron spectrum of the M\"uller model
in the HTSC theory. //
Acta Physica Polonica~A, 1993, vol.~84, No.~2, p.~293--313.
\bibitem{9} Izyumov~Yu.A., Letfulov~B.M.
Spin fluctuations and superconducting states in the Hubbard model
with a strong Coulomb repulsion. //
J.~Phys. - Condens. Matter, 1991, No.~3, \mbox{p.~5373.}
\bibitem{10} Izyumov~Yu.A., Letfulov~B.M.
Diagram technique for Hubbard operators.
Phase diagram for ($t-J$)-model. Preprint USSR Acad. Sci.
Ural Div., 90/2, 1990, \mbox{72~p.}
\bibitem{11} Stasyuk~I.V., Shvaika~A.M.
Dielectric instability and local
anharmonic model in the theory of high-$T_{\rm c}$ superconductivity. //
Physica~C, 1994, vol.~235--240, p.~2173--2174.
\bibitem{12} Stasyuk~I.V., Havrylyuk~Yu.
Phase transitions in pseudospin-electron model with direct
interaction between pseudospins. Preprint of the Institute for Condensed
Matter Physics, ICMP-98-18E, Lviv, 1998, \mbox{20~p.}
\bibitem{13} Stasyuk~I.V., Shvaika~A.M.
Dielectric instability and vibronic-type spectrum of local
anharmonic model of high-$T_{\rm c}$ superconductors. //
Ferroelectrics, 1997, vol.~192, \mbox{p.~1--10.}
\bibitem{14} Stasyuk~I.V., Shvaika~A.M. Pseudospin-electron model in
large dimensions. Preprint of the Institute for Condensed
Matter Physics, ICMP--98--20E, Lviv, 1998, \mbox{16~p.}
\bibitem{15} Stasyuk~I.V., Trachenko~K.O.
Investigation of locally anharmonic models of structural phase
transitions. Seminumerical approach. //
Cond. Matt. Phys., 1997, No.~9, \mbox{p.~89--107.}
\bibitem{16} Danyliv O.D.
Phase transitions in the two-sublattice pseudospin-electron
model of high temperature superconducting systems. //
Physica~C, 1998, vol.~309, \mbox{p.~303--314.}
\bibitem{17} Gervais~F.
Oxygen polarizability in ferroelectrics. A clue to understanding
superconductivity in oxides? // Ferroelectrics, 1992, vol.~130,
\mbox{p.~45--76.}
\bibitem{18} Kurtz~S.K., Hardy~J.R., Floken~J.W.
Structural phase transition,
ferroelectricity and high-temperature superconductivity. //
Ferroelectrics, 1988, vol.~87, \mbox{p.~29--40.}
\end{thebibliography}
\end{document}